\documentclass[aps, prl,english,preprintnumbers,twocolumn,floatfix,nofootinbib]{revtex4-1}
\usepackage{amssymb}
\usepackage[T1]{fontenc}
\usepackage{latexsym,epsfig}
\usepackage[latin9]{inputenc}
\usepackage{graphicx}
\usepackage{amsfonts}
\usepackage{epsfig}
\usepackage{dcolumn}
\usepackage{bm}
\usepackage{babel}
\usepackage{hyperref}
\usepackage{amsmath,mathrsfs}

\begin{document}

\date{\today }
\title{Non-equilibrium critical dynamics of the two-dimensional
Ashkin-Teller model at the Baxter line}
\author{H. A. Fernandes$^{1}$, R. da Silva$^{2}$, A. A. Caparica$^{3}$, J.
R. Drugowich de Felício$^{4}$}

\begin{abstract}
We investigate the short-time universal behavior of the two dimensional
Ashkin-Teller model at the Baxter line by performing time-dependent Monte
Carlo Simulations. First, as preparatory results, we obtain the critical
parameters by searching the optimal power law decay of the magnetization.
Thus, the dynamic critical exponents $\theta _{m}$ and $\theta _{p}$,
related to the magnetic and electric order parameters, as well as the
persistence exponent $\theta _{g}$, are estimated using heat-bath Monte
Carlo simulations. In addition, we estimate the dynamic exponent $z$ and the
static critical exponents $\beta $ and $\nu $ for both order parameters. We
propose a refined method to estimate the static exponents that considers two
different averages: one that combines an internal average using several
seeds with another which is taken over geographic variations in the power
laws. Moreover, we also performed the bootstrapping method for a
complementary analysis. Our results show that the ratio $\beta /\nu $
exhibits universal behavior along the critical line corroborating the
conjecture for both magnetization and polarization.
\end{abstract}

\maketitle

\affiliation{
$^{1}$Universidade Federal de Goi\'{a}s - UFG, Campus Jata\'{\i}, , Jata\'{\i}-GO, 78000-000, Brazil, \\
$^{2}$Instituto de F\'{\i}sica, Universidade Federal do Rio Grande do Sul, UFRGS,  Porto Alegre - RS, 91501-970, Brazil,\\
$^{3}$Instituto de F\'{\i}sica, Universidade Federal de Goi\'{a}s, Goi\^{a}nia-GO, 74.690-900, Brazil,\\
$^{4}$Departamento de F\'{\i}sica, Universidade de S\~{a}o Paulo,  Ribeir\~{a}o Preto-SP, 14040-901, Brazil}







\section{Introduction}

In 1971 Baxter \cite{Baxter1971} calculated the free energy of the symmetric
eight-vertex model and found out for the first time a continuous dependence
of the critical exponents on the coupling coefficients of the model. This
result seemed, in principle, to contradict the universality hypothesis \cite%
{Widom1965,Fisher1966,Kadanoff1967} which suggests that the critical
exponents should be constant and a variation would be possible only in the
case of a change in the symmetry. Despite this apparent contradiction,
Kadanoff and Wegner \cite{KadanoffWegner1971} and Wu \cite{Wu1971}, showed
independently a connection between the continuous variation of those
exponents and the presence of a marginal operator in the Hamiltonian by
demonstrating the equivalence of this model with an Ising model in a square
lattice without field. In this formulation, besides the interactions between
next-nearest-neighbors, there is still a four-body interaction and the
Hamiltonian is written as \cite{Baxter1972a}: 
\begin{eqnarray}
\beta \mathcal{H}_{8V} &=&-J_{1}\sum_{i,j=1}^{L}\sigma _{i,j+1}\sigma
_{i+1,j}-J_{2}\sum_{i,j=1}^{L}\sigma _{ij}\sigma _{i+1,j+1}-  \notag \\
&&\lambda \sum_{i,j=1}^{L}\sigma _{ij}\sigma _{i,j+1}\sigma _{i+1,j}\sigma
_{i+1,j+1},  \label{h8v}
\end{eqnarray}%
where $\sigma _{ij}=\pm 1$\ is the Ising spin at the site $(i,j)$\ of the
lattice, $\beta =(k_{B}T)^{-1}$, $k_{B}$\ and $T$\ being respectively the
Boltzmann constant and the temperature of the system. The sums run over all
spins and periodic boundary conditions are assumed: $\sigma _{L+1,j}=\sigma
_{1,j}$\ and $\sigma _{i,L+1}=\sigma _{i,1}$. The spins are coupled by the
coefficient $J_{1}$ in one direction and by $J_{2}$ in the other one and the
coefficient $\lambda $ couples four spins.

The symmetric eight-vertex model, also known as Baxter model, has only one
critical line, where $J_{1}=J_{2}=J$. This line is given by the equation 
\cite{Baxter1972a} 
\begin{equation}
\exp(-2\lambda)=\sinh(2J).  \label{critbax}
\end{equation}

Besides the eight-vertex model there are other models that exhibit
nonuniversality, e.g. the Ising model with competing interactions \cite%
{Barber1979} and the Ashkin-Teller model \cite{AshkinTeller1943}. The latter
was introduced in 1943 to describe a four-component system with
nearest-neighbors interactions, displaced on a two-dimensional lattice. Soon
after the Baxter's work, Fan \cite{Fan1972a} showed that the Ashkin-Teller
(AT) model could be represented by two superposed Ising systems and coupled
by a four-body interaction coefficient.

In this representation the Hamiltonian for the AT model is given by
two-species model: 
\begin{equation}
\begin{array}{lll}
\beta \mathcal{H}_{AT} & = & -K_{1}\sum_{i,j=1}^{L}\sigma _{i,j}(\sigma
_{i,j+1}+\sigma _{i+1,j}) \\ 
&  &  \\ 
&  & -K_{2}\sum_{i,j=1}^{L}\mu _{i,j}(\mu _{i,j+1}+\mu _{i+1,j}) \\ 
&  &  \\ 
&  & -K_{4}\sum_{i,j=1}^{L}\sigma _{i,j}\mu _{i,j}(\sigma _{i+1,j}\mu
_{i+1,j}+\sigma _{i,j+1}\mu _{i,j+1})\text{,}%
\end{array}
\label{hat}
\end{equation}%
where $\sigma _{i,j}=\pm 1$ ($\mu _{i,j}=\pm 1$) is the Ising spin at the
site ($i,j$) of the sublattice $\sigma $ ($\mu $), $K_{1}$ ($K_{2}$) is the
coupling coefficient of the spin variable $\sigma _{i,j}$ ($\mu _{i,j}$),
and $K_{4}$ is the four-body coefficient which couples the two Ising
systems. The sums run over all spins and periodic boundary conditions are
assumed: $\sigma (\mu )_{L+1,j}=\sigma (\mu )_{1,j}$\ and $\sigma (\mu
)_{i,L+1}=\sigma (\mu )_{i,1}$.

Wegner \cite{Wegner1972} showed that carrying out a duality transformation
in one of the lattices ($\mu $, for example), one can map the AT model into
a staggered eight-vertex model. This alternation does not disappears even
for the isotropic model ($K_{1}=K_{2}=K$) except at the self-dual line 
\begin{equation}
\exp (-2K_{4})=\sinh (2K),  \label{critAT}
\end{equation}%
where the AT model becomes equivalent to an isotropic eight-vertex model
with four-spin coupling constant ($\lambda $) given by 
\begin{equation}
\tanh (2\lambda )=\frac{\tanh (2K_{4})}{\tanh (2K_{4})-1}.  \label{duality}
\end{equation}%
which is critical if $K_{4}<\frac{1}{4}\ln 3$ \cite{Baxter1972a}, with
critical exponents related by \cite%
{Kadanoff1977,Kadanoff1979,DrugoKoberle1982}: 
\begin{equation}
2-\frac{1}{\nu _{AT}}=\frac{1}{(2-1/\nu _{8V})},  \label{Eq.Kadanoff}
\end{equation}%
where 
\begin{equation}
\frac{1}{\nu _{8V}}=1-\frac{2}{\pi }\sin ^{-1}(\tanh (2\lambda )).
\label{Eq:Kadanoff2}
\end{equation}

In the Ashkin-Teller model, besides the magnetization $M$ of each
sublattice, another order parameter is present: the polarization $P$. These
order parameters are defined as 
\begin{equation}
\begin{array}{cccccc}
M_{\sigma } & = & \frac{1}{L^{2}}\left\langle \sum_{i,j=1}^{L}\sigma
_{i,j}\right\rangle , & M_{\mu } & = & \frac{1}{L^{2}}\left\langle
\sum_{i,j=1}^{L}\mu _{i,j}\right\rangle , \\ 
&  &  &  &  &  \\ 
P & = & \frac{1}{L^{2}}\left\langle \sum_{i,j=1}^{L}\sigma _{i,j}\mu
_{i,j}\right\rangle  &  &  & 
\end{array}
\label{orderparameters}
\end{equation}%
where $\left\langle \cdot \right\rangle $\ denotes the ensemble average: 
\begin{equation*}
\left\langle (\cdot )\right\rangle =\frac{1}{Z}\sum_{\left\{ \sigma
_{i,j},\mu _{i,j}\right\} }(\cdot )\exp \left[ -\beta \mathcal{H}%
_{AT}(\left\{ \sigma _{i,j},\mu _{i,j}\right\} _{i,j=1}^{L})\right] 
\end{equation*}%
with $Z=\sum_{\left\{ \sigma _{i,j},\mu _{i,j}\right\} }\exp \left[ -\beta 
\mathcal{H}_{AT}(\left\{ \sigma _{i,j},\mu _{i,j}\right\} _{i,j=1}^{L})%
\right] $.

However, as we are dealing with the isotropic version of the model, the
spins of each sublattice are symmetric and, in this case, their
magnetizations will have the same behavior. Then, the net result is that the
number of samples for the magnetization is doubled. Henceforth, we consider
only two order parameters: the magnetization $(M)$ (that includes both
sublattices) and the polarization $(P)$.

The purpose of this paper is to study the dynamic critical behavior of the
Ashkin-Teller model to obtain the dynamic exponents $\theta _{g}$, $\theta $%
, and $z$, as well as the static exponents $\beta $ and $\nu $ for both
order parameters. To reach our goal, we carry out short-time Monte Carlo
simulations in the two-dimensional isotropic AT model by considering the
duality relation between these two models, Eq. (\ref{duality}). The paper is
organized as follows. In the next section, we briefly present the
non-equilibrium technique as well as the scaling relations used in this
work. In the third section, we find out the critical exponents of the
Ashkin-Teller model. Finally, in the fourth section we present our
conclusions.

\section{Critical dynamics for the model}

Until a few years ago, it was a common sense that no universal behavior
could be found in systems during the initial stage of the relaxation
process. As a result, critical properties of these systems, like transition
temperatures and critical exponents, were obtained only in equilibrium. The
numerical calculation of such values was not a simple task, due to the
severe \textit{critical slowing down} which takes place in the vicinity of
the criticality. Many efforts have been endued to circumvent this
difficulty, for instance, the cluster algorithm \cite%
{SwendsenWang1987,Wolff1989} has proven to be very efficient in the study of
static properties of systems. Nevertheless, in that case the original
dynamic class of universality is violated, leading to normally small values
for the dynamic critical exponents. Another way to avoid problems with the 
\textit{critical slowing down} was proposed by Janssen \textit{et al.} \cite%
{Janssen1989} and Huse \cite{Huse1989}. Using renormalization group
techniques and numerical calculation, respectively, they showed that the
critical relaxation of a system initially at very high temperature exhibits
universality and scaling behavior even in the initial steps of evolution.
The so-called short-time regime became therefore an important method in the
study of phase transitions and critical phenomena.

The dynamic scaling relation obtained by Janssen \textit{et al.} for the 
\textit{k}-th moment of the magnetization, extended to systems of finite
size \cite{Li1995,Zheng1998}, is written as 
\begin{equation}
\overline{M^{k}}(t,\tau ,L,m_{0})=b^{-\frac{k\beta }{\nu }}\overline{M^{k}}%
(b^{-z}t,b^{\frac{1}{\nu }}\tau ,b^{-1}L,b^{x_{0}}m_{0}).  \label{eq1}
\end{equation}%
Here $t$ is the time evolution, $b$ is an arbitrary spatial rescaling
factor, $\tau =\left( T-T_{c}\right) /T_{c}$ is the reduced temperature and $%
L$ is the linear size of the square lattice. This evolution is governed by a
new dynamic exponent $\theta $ independent of the well known static critical
exponents and the dynamic exponent $z$. This new exponent characterizes the
so-called \textit{critical initial slip}, the anomalous behavior of the
magnetization when the system is quenched to the critical temperature $T_{c}$%
. In addition, a new critical exponent $x_{0}$ which represents the
anomalous dimension of the initial magnetization $m_{0}$, is introduced to
describe the dependence of the scaling behavior on the initial conditions.
This exponent is related to $\theta $ as $x_{0}=\theta z+\beta /\nu $.

From Eq. (\ref{eq1}), the scaling relations for the \textit{k}-th moment of
the magnetization and polatization of the Ashkin-Teller model are given,
respectively, by 
\begin{equation}
\overline{M^{k}}(t,\tau ,L,m_{0})=b^{-\frac{k\beta _{m}}{\nu }}\overline{%
M^{k}}(b^{-z_{m}}t,b^{\frac{1}{\nu }}\tau ,b^{-1}L,b^{x_{m}}m_{0})
\label{eq2}
\end{equation}%
and 
\begin{equation}
\overline{P^{k}}(t,\tau ,L,p_{0})=b^{-\frac{k\beta _{p}}{\nu }}\overline{%
P^{k}}(b^{-z_{p}}t,b^{\frac{1}{\nu }}\tau ,b^{-1}L,b^{x_{p}}p_{0}),
\label{eq3}
\end{equation}%
where $p_{0}$ is the initial polarization of the system. Here, differently
from $\left\langle O\right\rangle $, the average $\overline{O}$\ describes
an average over different random evolutions and over initial conditions of
the system.

In this work the dynamic critical exponents $\theta _{m}$ and $\theta _{p}$
are obtained through two different approaches, a time correlation of the
magnetization \cite{Tania1998} 
\begin{equation}
Q_{M}(t)=\overline{M(0)M(t)}\sim t^{\theta _{m}}  \label{theta}
\end{equation}%
and 
\begin{equation}
Q_{P}(t)=\overline{P(0)P(t)}\sim t^{\theta _{p}},  \label{theta-1}
\end{equation}%
and the scaling forms 
\begin{equation}
\overline{M}(t)\sim m_{0}t^{\theta _{m}}  \label{thtm}
\end{equation}%
and 
\begin{equation}
\overline{P}(t)\sim p_{0}t^{\theta _{p}}.  \label{thtp}
\end{equation}

In order to see such power law behaviors we can look into some details of
scaling relation. Taking into account the magnetization (we have a similar
analysis for the polarization), after the scaling $b^{-1}L=1$\ at the
critical temperature $T=$\ $T_{c}$, the first ($k=1$) moment of the order
parameter is $\overline{M}(t,L,m_{0})=L^{-\beta /\nu }\overline{M}%
(L^{-z}t,L^{x_{0}}m_{0})$. Denoting $u=tL^{-z}$\ and $w=L^{x_{0}}m_{0}$, one
has $\overline{M}(u,w)=L^{-\beta /\nu }\overline{M}(L^{-z}t,L^{x_{0}}m_{0})$%
. Hence, the derivative with respect to $L$\ is given by

\begin{eqnarray*}
\partial _{L}\overline{M} &=&(-\beta /\nu )L^{-\beta /\nu -1}\overline{M}%
(u,w) \\
&&+L^{-\beta /\nu }[\partial _{u}\overline{M}\partial _{L}u+\partial _{w}%
\overline{M}\partial _{L}w]\text{,}
\end{eqnarray*}%
where one has explicitly $\partial _{L}u=-ztL^{-z-1}$\ and $\partial
_{L}w=x_{0}m_{0}L^{x_{0}-1}$. In the limit $L\rightarrow \infty $, which
implicates in $\partial _{L}\overline{M}\rightarrow 0$, one has $%
x_{0}w\partial _{w}\overline{M}-zu\partial _{u}\overline{M}-\beta /\nu 
\overline{M}=0$. The separability of the variables $u$\ and $w$, i.e., $%
\overline{M}(u,w)=M_{u}(u)M_{w}(w)$\ leads to%
\begin{equation*}
x_{0}wM_{w}^{\prime }/M_{w}=\beta /\nu +zuM_{u}^{\prime }/M_{u}\text{,}
\end{equation*}%
where the prime means the derivative with respect to the argument. Since the
left-hand side of this equation depends only on $w$ and the right-hand side
depends only on $u$, both sides must be equal to a constant $c$. Thus, $%
M_{u}(u)=u^{c/z}-\beta /(\nu z)$ and $M_{w}(w)=w^{c/x_{0}}$, resulting in $%
\overline{M}(u,w)=m_{0}^{c/x_{0}}L^{\beta /\nu }t^{(c-\beta /\nu )/z}$.
Returning to the original variables, one has 
\begin{equation}
\overline{M}(t,L,m_{0})=m_{0}^{c/x_{0}}t^{(c-\beta /\nu )/z}\mathbf{.}
\label{Eq:main_criticality}
\end{equation}

By choosing $c=x_{0}$\ at criticality ($\tau =0$), one obtains $\overline{M}%
_{m_{0}}\sim m_{0}t^{\theta }$, as previously reported in Eq. (\ref{thtm})
and (\ref{thtp}), where $\theta =(x_{0}-\beta /\nu )/z$.\ This corresponds
to a regime of small initial magnetization soon after a finite time scaling $%
b=t^{1/z}$\ in Eq.~(\ref{eq1}). We therefore obtain $\overline{M}%
(t,m_{0})=t^{-\beta /(\nu z)}\overline{M}(1,t^{x_{0}/z}m_{0})$. By calling $%
x=t^{x_{0}/z}m_{0}$, an expansion of the averaged magnetization around $x=0$%
\ results in $\overline{M}(1,x)=\overline{M}(1,0)+\left. \partial _{x}%
\overline{M}\right\vert _{x=0}x+O(x^{2})$. By construction $\overline{M}%
(1,0)=0$\ and, since $u=t^{x_{0}/z}m_{0}\ll 1$, we can discard quadratic
terms resulting similarly in $\langle M\rangle _{m_{0}}\sim m_{0}t^{\theta }$%
. This anomalous behavior of initial magnetization is valid only for a
characteristic time scale $t_{\max }$\ $\sim m_{0}^{-z/x_{0}}$.

Another dynamic critical exponent is obtained far from equilibrium by
following the behavior of the global persistence probability $G(t)$ \cite%
{Majumdar1996}, the probability of the order parameter does not change its
sign up to the time $t$. For the magnetization and polarization, it decays
respectively as 
\begin{equation}
G_{M}(t)\sim t^{-\theta _{g_{m}}},  \label{perm}
\end{equation}%
and 
\begin{equation}
G_{P}(t)\sim t^{-\theta _{g_{p}}},  \label{perp}
\end{equation}%
where the exponents $\theta _{g_{m}}$ and $\theta _{g_{p}}$ are the global
persistence exponents of the magnetization and polarization, respectively.

As pointed out in Ref. \cite{Majumdar1996} and shown in several works \cite%
{Majumdar2003,Schulke1997,Oerding1997,Dasilva2003,Dasilva2005,Ren2003,Albano2001,Saharay2003,Hinrichsen1998,Sen2004,Zheng2002,Fernandes2006a,Fernandes2006b,Fernandes2006c}%
, the global persistence exponent is an independent critical index and is
closely related to the non-Markovian character of the process. On the
contrary, if the process would be a Markovian one, this exponent should obey
the equation 
\begin{equation}
\theta_g z=-\theta z+\frac{d}{z}-\frac{\beta}{\nu}.
\end{equation}

The dynamic critical exponents $z_{m}$ and $z_{p}$ are obtained using the
ratios \cite{Silva20021} 
\begin{equation}
F_{2_{M}}(t)=\frac{\overline{M(t)^{2}}_{m_{0}=0}}{\overline{M(t)}%
_{m_{0}=1}^{2}}\sim t^{d/z_{m}}  \label{f2m}
\end{equation}%
and 
\begin{equation}
F_{2_{P}}(t)=\frac{\overline{P(t)^{2}}_{p_{0}=0}}{\overline{P(t)}%
_{p_{0}=1}^{2}}\sim t^{d/z_{p}},  \label{f2p}
\end{equation}%
where $d$\ is the dimension of the system and the average is over different
samples with initial states $m_{0}$\ and $p_{0}$, respectively.

The first moment of the magnetization in Eq. (\ref{f2m}) (the denominator)
is obtained by making $c=0$ in Eq. (\ref{Eq:main_criticality}) and
considering which that such power law decays from ordered initial state ($%
m_{0}=1$). Since the system has no dependence on initial conditions, one has 
\begin{equation}
\overline{M}_{m_{0=1}}(t)\sim t^{-\frac{\beta _{m}}{\nu _{m}z_{m}}}
\label{Eq:Decay_mag}
\end{equation}%
The same analysis can be done for the polarization obtaining 
\begin{equation}
\overline{P}_{p_{0}=1}(t)\sim t^{-\frac{\beta _{p}}{\nu _{p}z_{p}}}.
\label{Eq:Decay_pol}
\end{equation}

On the other hand, the second moment of the magnetization in Eq. (\ref{f2m})
(the numerator) can be written as

\begin{equation*}
\overline{M_{m_{0}=0}^{2}}=\frac{1}{L^{2d}}\sum\limits_{i=1}^{L^{d}}\left%
\langle \sigma _{i}^{2}\right\rangle +\frac{1}{L^{2d}}\sum%
\limits_{i}^{L^{d}}\left\langle \sigma _{i}\sigma _{j}\right\rangle \approx
L^{-d}
\end{equation*}%
for a fixed $t$. By taking into account $k=2$ in Eq. (\ref{eq2}) with $%
b=t^{1/z_{m}}$ and considering that the spin-spin correlation $\left\langle
\sigma _{i}\sigma _{j}\right\rangle $ is negligible form $m_{0}=0$, we obtain

\begin{equation}
\begin{array}{lll}
\overline{M_{m_{0}=0}^{2}}(t,L) & \approx & t^{\frac{-2\beta _{m}}{\nu
_{m}z_{m}}}\overline{M_{m_{0}=0}^{2}}(1,bL) \\ 
\  & \  & \  \\ 
\  & = & t^{\frac{-2\beta _{m}}{\nu _{m}z_{m}}}(bL)^{-d} \\ 
\  & \  & \  \\ 
\  & \sim & t^{(d-\frac{2\beta _{m}}{\nu _{m}})/z_{m}}%
\end{array}
\label{M2}
\end{equation}%
and similarly,

\begin{equation}
\overline{P_{p_{0}=0}^{2}}(t,L)\sim t^{(d-\frac{2\beta _{p}}{\nu _{p}}%
)/z_{p}}  \label{P2}
\end{equation}%
for the second moment of the polarization. Therefore, the power laws given
by Eqs. (\ref{f2m}) and (\ref{f2p}) can be easily verified.

This approximation proved to be very efficient in estimating the exponent $z$%
, according to results for the Ising model, the $q=3$ and $q=4$ Potts models 
\cite{Silva20021}, the tricritical point of the Blume-Capel model \cite%
{Silva20022}, metamagnetic model \cite{Silva2013}, ANNNI model \cite%
{Silva2013-ANNI}, spin models based on generalized Tsallis statistics \cite%
{Silva2012-Tsallis}, Z5 model \cite{Silva2014-Z5}, the Baxter-Wu model \cite%
{Arashiro2003}, the double-exchange model \cite{Fernandes2005}, Heisenberg
model \cite{Fernandes2006a} and even models without defined Hamiltonian
(see, for example, Refs. \cite{Dasilva2005,Rdasilva-contact2004}).

The static exponents must be obtained via other power laws. When $%
L\rightarrow \infty $, one has $\overline{M}(t,\tau )=b^{-k\beta /\nu }%
\overline{M}(b^{-z}t,b^{1/\nu }\tau )$. By scaling $b^{-z}t=1$, we have $%
\overline{M}(t,\tau )=t^{-\beta /(\nu z)}f(t^{1/(\nu z)}\tau )$\ where $%
f(x)= $\ $\overline{M}(1,x)$\ and so $\partial \ln \overline{M}(t,\tau
)/\partial \tau =\frac{1}{\langle M\rangle }\frac{\partial }{\partial \tau }%
\overline{M}=t^{1/(\nu z)}f(t^{1/(\nu z)}\tau )$. Therefore we have 
\begin{equation}
D_{M}(t)=\left. \frac{\partial \ln \overline{M}}{\partial \tau }\right\vert
_{\tau =0}=f_{0}\cdot t^{1/(\nu _{m}z_{m})}\sim t^{\phi _{m}}  \label{1niz}
\end{equation}%
where $f_{0}=f(0)$\ is a constant and $\phi _{m}=$\ $1/(\nu _{m}z_{m})$.
Since we have already estimated the exponent $z_{m}$\ (Eq. (\ref{f2m})), we
are able to obtain $\nu _{m}$. With these two exponents in hand, we can
obtain $\beta _{m}$\ by estimating the exponent $\mu _{m}=\beta _{m}/(\nu
_{m}z_{m})$\ from Eq. (\ref{Eq:Decay_mag}). By changing $M$\ by $P$, we have 
\begin{equation}
D_{P}(t)=\left. \frac{\partial \ln \overline{P}}{\partial \tau }\right\vert
_{\tau =0}\sim t^{\phi _{p}}  \label{1nizp}
\end{equation}%
\ with $\phi _{p}=$\ $1/(\nu _{p}z_{p})$, where the exponents $\nu _{p}$ and 
$\beta _{p}$ are obtained by following the same procedures adopted for the
magnetization.

\section{Some details about heat-bath Monte Carlo simulations}

\label{Section:Details_Monte_Carlo}

In this section, we describe with some details how the heat-bath Monte Carlo
simulations are carried out in our work to evolve the spins. The interesting
point here is that the transition occurs for the pair of spins $\left(
\sigma _{i,j},\ \mu _{i,j}\right) $\ and not for single spins since we have
coupled lattices. Moreover, this transition does not depend on current spin.
So, the transition probabilities of each possible pair: $(+,+)$, $(-,+)$, $%
(+,-)$, and $(-,-)$\ are calulated by%
\begin{equation*}
p\left[ \cdot \rightarrow \left( \sigma _{i,j},\ \mu _{i,j}\right) \right] =%
\frac{1}{S}\exp \left[ -E\left( \sigma _{i,j},\ \mu _{i,j}\right) \right]
\end{equation*}%
with $S=e^{-E(+,+)}+e^{-E(-,+)}+e^{-E(+,-)}+e^{-E(-,-)}$\ and 
\begin{equation*}
\begin{array}{ccc}
E(+,+) & = & -K\left( \Delta _{i,j}+\Psi _{i,j}\right) -K_{4}\Phi _{i,j}%
\text{,} \\ 
&  &  \\ 
E(-,+) & = & -K\left( \Psi _{i,j}-\Delta _{i,j}\right) +K_{4}\Phi _{i,j}%
\text{,} \\ 
&  &  \\ 
E(+,-) & = & -K\left( \Delta _{i,j}-\Psi _{i,j}\right) +K_{4}\Phi _{i,j}%
\text{,} \\ 
&  &  \\ 
E(-,-) & = & K\left( \Delta _{i,j}+\Psi _{i,j}\right) -K_{4}\Phi _{i,j}\text{%
,}%
\end{array}%
\end{equation*}%
where 
\begin{equation*}
\Delta _{i,j}=\sigma _{i+1,j}+\sigma _{i-1,j}+\sigma _{i,j+1}+\sigma _{i,j-1}%
\text{,}
\end{equation*}%
\bigskip 
\begin{equation*}
\Psi _{i,j}=\mu _{i+1,j}+\mu _{i-1,j}+\mu _{i,j+1}+\mu _{i,j-1}
\end{equation*}%
and 
\begin{equation*}
\Phi _{ij}=\sigma _{i+1,j}\mu _{i+1,j}+\sigma _{i-1,j}\mu _{i-1,j}+\sigma
_{i,j+1}\mu _{i,j+1}+\sigma _{i,j-1}\mu _{i,j-1}\text{.}
\end{equation*}

For the AT model the order parameters correspond to time-dependent
magnetization, polarization, as well as, their superior moments, here
represented by a general symbol $O$ defined via our MC simulations as an
average over all $L^{2}\ $spins and over the different $N_{run}$\ runs
(different time evolutions):

\begin{equation}
\overline{O}(t)=\frac{1}{N_{run}\ L^{2}}\sum\limits_{k=1}^{N_{run}}\sum%
\limits_{i,j=1}^{L}O_{i,j,k}(t)\text{,}  \label{magnetizacao}
\end{equation}%
where the index $k=1,...,N_{s}$\ denotes the corresponding run of each
simulation. The ordered state is ferromagnetic, with all (or most of) the
spins pointing either up or down.

As discussed in the previous section, the lattice's initial condition to be
simulated in our study depends on the scaling relation as follows:

a) Eqs. (\ref{theta}) and (\ref{theta-1}): To obtain such power laws, the
averages are obtained from a set of runs with initially random
configurations allowing the direct calculation of the dynamic exponents $%
\theta _{m}$\ and $\theta _{p}$. Here, the only requirement is that $\langle
m_{0}\rangle =\langle p_{0}\rangle =0$. Unfortunately, the huge fluctuations
for $P(t)$, even for $N_{run}=300,000$\ runs, prevented us of estimating $%
\theta _{p}$\ through this method.

b) Eqs. (\ref{thtm}) and (\ref{thtp}): In order to obtain the same exponents 
$\theta _{m}$\ and $\theta _{p}$ we use these alternative equations.
However, in this case, a careful preparation of the initial order parameters
($m_{0}$\ and $p_{0}$) is needed, besides the limit procedures $%
m_{0}\rightarrow 0$\ and $p_{0}\rightarrow 0$. Here we used $N_{run}=100,000$%
\ runs.

c) Eqs. (\ref{Eq:Decay_mag}) and (\ref{Eq:Decay_pol}): In order to perform
the simulations to obtain the exponents by these power laws, we used ordered
initial states, which means $m_{0}=1$\ and $p_{0}=1$. In this particular
case the simulations do not present sensitive fluctuations and for all cases
we used $N_{run}=4,000$\ runs.

d) Eqs. (\ref{M2}) and (\ref{P2}): When computing the second moment of the
magnetization or polarization, we used $m_{0}=0$\ (half (randomly choosen)
of spins up and the other half of the spins down) and $N_{run}=4,000$\ runs.

e) Eq. (\ref{1niz}): When dealing with Monte Carlo simulations, the partial
derivative is approximated in first order by the difference 
\begin{equation}
\left. \frac{\partial \ln \left. \overline{M}(t,\tau )\right\vert _{m_{0}=1}%
}{\partial \tau }\right\vert _{\tau =0}\approx \frac{1}{2\varepsilon }\ln %
\left[ \frac{\left. \overline{M}(t,T_{c}+\varepsilon )\right\vert _{m_{0}=1}%
}{\left. \overline{M}(t,T_{c}-\varepsilon )\right\vert _{m_{0}=1}}\right]
\label{aproximation}
\end{equation}%
where $\varepsilon <<1$. It is clear from Eq. (\ref{aproximation}) that two
independent simulations are necessary to obtain the exponent $1/\nu z$: one
of them evolves at the temperature $T_{c}+\varepsilon $, and the other one
evolves at $T_{c}-\varepsilon $. Here we used $N_{run}=4,000$\ runs for $%
\left. \overline{M}(t,T_{c}+\varepsilon )\right\vert _{m_{0}=1}$\ and $%
N_{run}=4,000$\ runs for $\left. \overline{M}(t,T_{c}-\varepsilon
)\right\vert _{m_{0}=1}$\ since we start from ordered initial states.

\section{Localization of critical points: Power law optimization}

In this section we performed some initial simulations to give more knowledge
about the criticality of the AT model. The theoretical predictions of the
critical line are described by 
\begin{equation}
K_{4}(K)=-\frac{1}{2}\ln (\sinh (2K))\text{,}  \label{Eq. Critical_line}
\end{equation}
therefore, let us consider a particular critical point of this curve,
denoted by $(K^{(c)},K_{4}^{(c)})$, which corresponds to a particular
critical coefficient $J_{c}$ of the Baxter model, such that $\lambda _{c}=- 
\frac{1}{2}\ln \sinh (2J_{c})$. Hence, we obtain $K_{4}^{(c)}=\frac{1}{2}%
\tanh ^{-1}\left( \frac{\tanh (2\lambda _{c})}{\tanh (2\lambda _{c})-1}%
\right) $ and $K^{(c)}=\frac{1}{2}\sinh ^{-1}(\exp (-2K_{4}^{(c)}))$.

So, the tangent line to the curve $K_{4}(K)=-\frac{1}{2}\ln (\sinh (2K))$\
passing by for the critical point $(K^{(c)},K_{4}^{(c)})$\ is written as: 
\begin{equation*}
K_{4}^{\Vert }=-\coth (2K_{c})(K-K_{c})-\frac{1}{2}\ln (\sinh (2K_{c}))\text{%
,}
\end{equation*}%
and the perpendicular line to this tangent line, can be written as:%
\begin{equation}
K_{4}^{\perp }=\tanh (2K_{c})(K-K_{c})-\frac{1}{2}\ln (\sinh (2K_{c}))\text{.%
}  \label{Eq:perpendicular}
\end{equation}

In Table \ref{table:points_revised} we collect the considered points $(J)$
as well as the four-body coupling constants $(\lambda )$ in the critical
line of the Baxter model. The corresponding coefficients $K$ and $K_{4}$ for
the Ashkin-Teller model, calculated from Eqs. (\ref{critbax}), (\ref{critAT}%
), and (\ref{duality}), are also shown in this table. We choose five points $%
J=0.4$, Ising Model which corresponds to $J=\frac{1}{2}\ln (1+\sqrt{2})$, $%
J=0.5$, $q=3$ Potts (TSP) model corresponding to $J=0.596...$ and $q=4$
Potts (FSP) model corresponding to $J\rightarrow +\infty $.

\begin{table*}[tbh]
\centering%
\begin{tabular}{ccccc}
\hline
$J$ & $\lambda $ & $K$ & $K_{4}$ & Critical Point \\ \hline
0.4 & 0.059332097 & 0.489889651 & -0.067369092 &  \\ 
$0.5\ln (1+\sqrt{2})$ & 0 & $0.5\ln (1+\sqrt{2})$ & 0 & Ising Model \\ 
0.5 & -0.080719681 & 0.393334281 & 0.069427372 &  \\ 
0.596439479 & -0.20159986 & 0.347625611 & 0.142089631 & TSP Model \\ 
+$\infty $ & -$\infty $ & $(\ln 3)/4$ & $(\ln 3)/4$ & FSP Model \\ \hline
\end{tabular}%
\caption{The five points in the self-dual critical line.}
\label{table:points_revised}
\end{table*}

\begin{figure}[h]
\begin{center}
\includegraphics[width=\columnwidth]{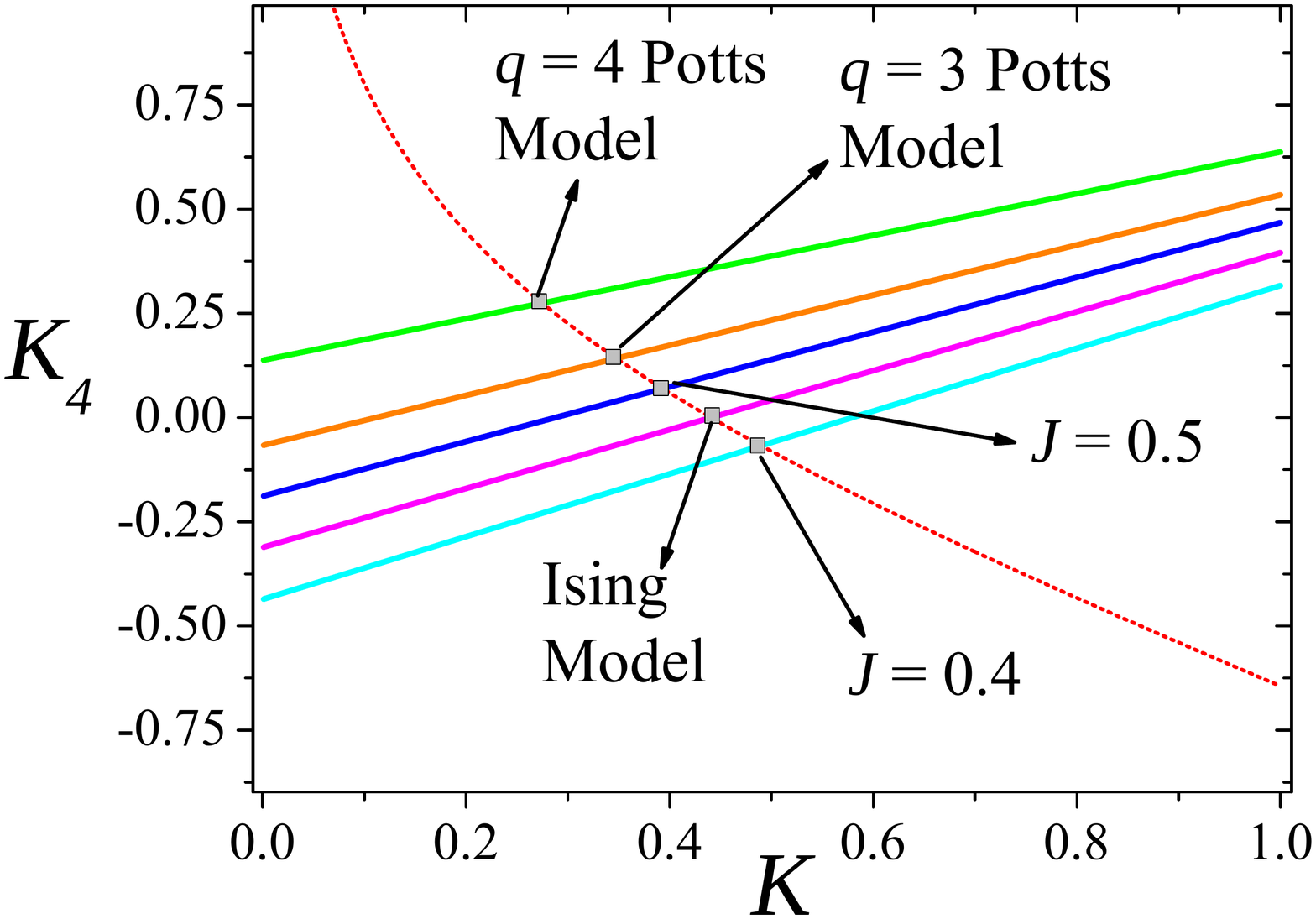}
\end{center}
\caption{Critical line described by equation $K_{4}=-\frac{1}{2}\ln (\sinh
(2K))$. The points correspond to $J=0.4$, $J=0.5$, Ising model, $q=3$ Potts
(TSP) model and $q=4$ Potts (FSP) model. The perpendicular lines passing
through each point are also presented. }
\label{Fig:points}
\end{figure}

In Figure \ref{Fig:points} we present the critical line of the AT model and
ilustrate the points to be considered in this study, as well as the
perpendicular lines for each point.\ 

Our initial plan was to study the phase transition points of the AT model
via time-dependent MC simulations by estimating the best $K$\ given as input
the parameter $K^{(\min )}$\ (inicial value) and run simulations for
different values of $K$\ according to a resolution $\Delta K$.

We performed this task for the five points in Fig. \ref{Fig:points} by
taking into account only the magnetization and the analysis was carried out
by using an approach developed in Ref. \cite{Silva2012-Tsallis} in the
context of generalized statistics. This tool had also been applied
successfully to study multicritical points, for example, tricritical points 
\cite{Silva2013} and Lifshitz point of the ANNNI model \cite{Silva2013-ANNI}%
, Z5 model \cite{Silva2014-Z5} and also in models without defined
Hamiltonian \cite{Silva-Fernandes2015}.

Since at criticality it is expected that the order parameter obeys the power
law behavior of Eq. (\ref{Eq:Decay_mag}), for each value $K=K^{(\min
)}+i\Delta K$, with $i=1,...,n$, where $n=\left\lfloor (K^{(\max )}-K^{(\min
)})/\Delta K\right\rfloor $, we performed MC simulations and calculated the
coefficient of determination, which is given by 
\begin{equation}
r=\frac{\sum\limits_{t=1}^{N_{MC}}(\left\langle \ln \overline{M}%
\right\rangle -a-b\ln t)^{2}}{\sum\limits_{t=1}^{N_{MC}}(\left\langle \ln 
\overline{M}\right\rangle -\ln \langle M\rangle (t))^{2}}\text{,}
\label{determination_coefficient}
\end{equation}%
with $\left\langle \ln \overline{M}\right\rangle
=(1/N_{MC})\sum\nolimits_{t=1}^{N_{MC}}\ln \overline{M}(t)$, and the
critical value $K_{c}\ $corresponds to $K^{(opt)}=\arg \max_{K\in \lbrack
K^{(\min )},K^{(\max )}]}\{r\}$. The coefficient $r$\ has a very simple
explanation: it measures the ratio: (expected variation)/(total variation).
The bigger the $r$, the better the linear fit in log-scale, and therefore,
the better the power law which corresponds to the critical parameter except
for an order of error $\Delta K$.

\begin{figure}[h]
\begin{center}
\includegraphics[width=\columnwidth]{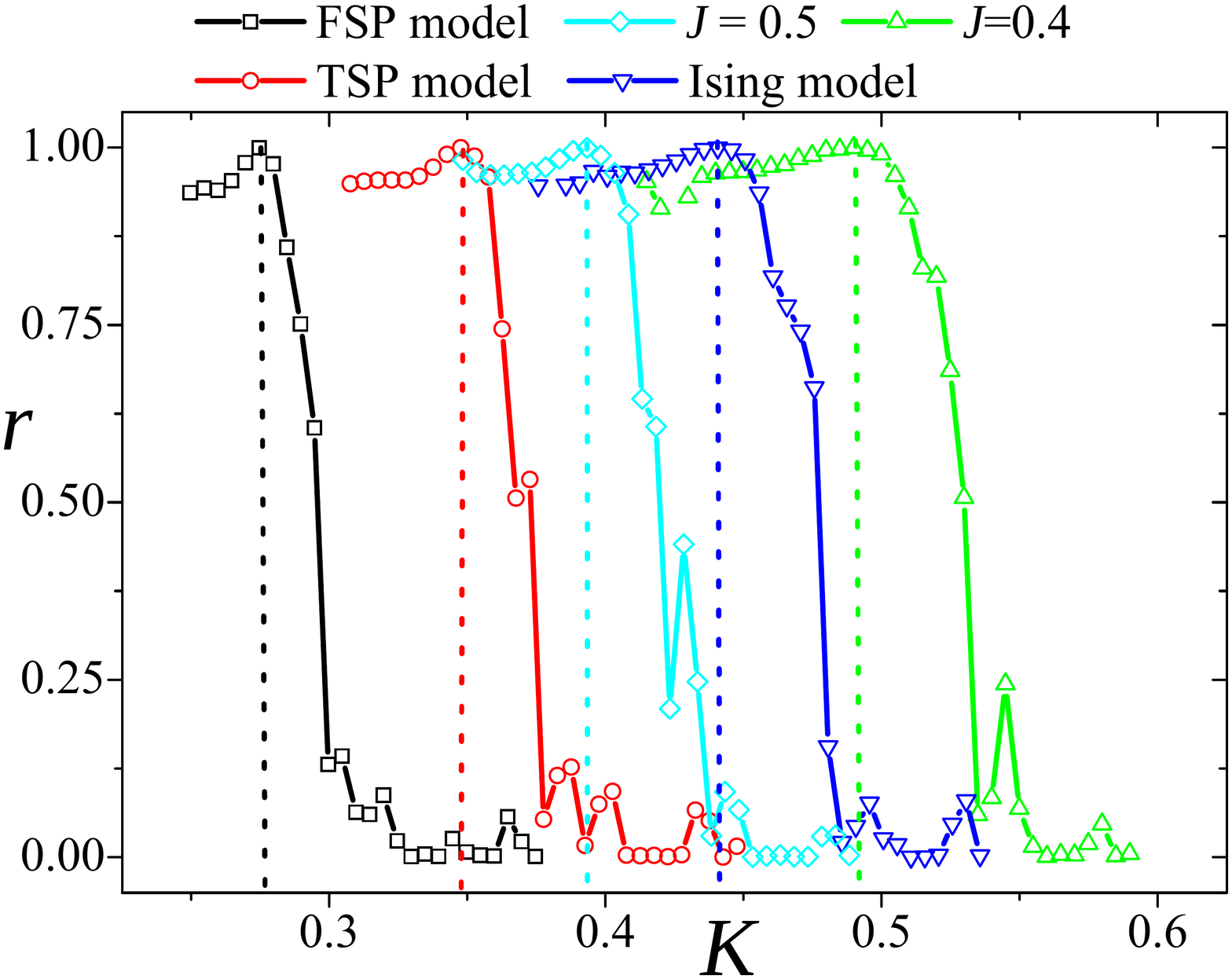}
\end{center}
\caption{Coefficient of determination $r$ as function of $K$ walking on the
perpendicular line presented in Fig. \protect\ref{Fig:points}. The maximum
occurs at the expected critical points. }
\label{Fig:Optimization}
\end{figure}

Particularly for these simulations, whose main aim is to check the critical
parameter, we used only $N_{MC}=300$\ MCsteps but, for the simulations used
to estimate the static critical exponents, we used $N_{MC}=1000$ MCsteps.

In Fig. \ref{Fig:Optimization} we can observe that the maximum occurs
exactly in the point ($K^{(c)},K_{4}^{(c)}$) as conjectured by Eq. (\ref{Eq.
Critical_line}) for each point. This figure shows $r$\ as function of $K$\
walking on the perpendicular line given by Eq. (\ref{Eq:perpendicular}).
Since we corroborate such conjecture using an optimizer based on MC
simulations, we are now prepared to study the critical exponents (dynamic
and static ones) for these points.

\section{Results}

In this article we study the short-time critical dynamics of the
Ashkin-Teller model \cite{Baxter1972a} by carring out Monte Carlo
simulations in five points (see Table \ref{table:points_revised}) along the
Baxter line where the model presents nonuniversal behavior.

We estimate the dynamic critical exponents $\theta _{g_{m}}$, $\theta
_{g_{p}}$, $\theta _{m}$, $\theta _{p}$, $z_{m}$, and $z_{p}$ for each
considered critical point as well as the static critical exponents: $\nu
_{m} $, $\nu _{p}$, $\beta _{m}$, and $\beta _{p}$. We elaborate a more
detailed statistical procedure to estimate the static exponents since their
sensitivity deserves more attention. Among the points we take into account,
we include the critical points of the Ising, TSP, and FSP models. The
exponents $\theta _{g_{m}}$, $\theta _{m}$, and $z_{m}$ were obtained
numerically for the critical points of the two-dimensional Ising model \cite%
{Zheng1998,Li1994,Li1996,Okano1997,Grassberger1995}, the TSP model \cite%
{Zheng1998,Silva20021,Schulke1995,Chatelain2004}, and the FSP model \cite%
{Fernandes2006b,Silva20021,Arashiro2003,Chatelain2004,Silva2004,Arashiro2009}%
. These last two exponents, as well as the exponents $\theta _{p}$ and $%
z_{p} $ were calculated for some points on the self-dual critical line of
the Ashkin-Teller model by Li \textit{et al.} \cite{Li1997}. In addition,
the exponents $z_{m}$ and $z_{p}$ were estimated for some points on the
critical line for the Baxter model by Takano \cite{Takano1996}. As far as we
know, the dynamic critical exponents $\theta _{g_{m}}$, and $\theta _{g_{p}}$
were not found yet for the Ashkin-Teller model. It is important to mention
that $\theta_{m}$ and $\theta_{p}$ have not yet been obtained by power law
correlations, as well as the exponents $z_{m}$ and $z_{p}$ which have not
yet been studied through the method that mixes initial conditions. Both
methods are employed in this paper. On the other hand, for the static
exponents, conjectures assert that the ratio $\beta _{m}/\nu _{m}=1/8$ for
the entire critical line while $\beta _{p}/\nu _{p}$ is not constant as $J$
increases. Hence, this fact deserves attention and a detailed study.

In our simulations we use square lattices of linear sizes $L=64$, $128$, and 
$256$\ and the system evolves in contact with a thermal bath in five points
on the self-dual critical line of the AT model. Our estimates for each
exponent and the corresponding error are obtained from five independent
seeds of $N_{run}$\ runs each one as previously described in Section \ref%
{Section:Details_Monte_Carlo}. However, since the two sublattices of the
model ($\sigma $\ and $\mu $) are symmetrical, the number of effective bins
for the magnetization are doubled. In order to measure the slopes of the
power laws described above (in double-log scale) we consider the time
interval [150,300] for the dynamic exponents. For the static ones, a more
detailed statistical tool was prepared taking into consideration averages
over different seeds and geographic variations. In this case the maximal
number of MC steps was $N_{MC}=10^{3}$.

\subsection{The dynamic critical exponents $\protect\theta _{g_{m}}$ and $%
\protect\theta _{g_{p}}$}

\label{Sec:exponents_theta_persistence}

The first exponents we calculate are the global persistence exponents $%
\theta _{g_{m}}$ and $\theta _{g_{p}}$ that are achieved when one considers
the global persistence probabilities $G_{M}(t)$ and $G_{P}(t)$, Eqs. (\ref%
{perm}) and (\ref{perp}), which are defined as the probabilities of the
order parameters (magnetization and polarization) not changing their signs
up to the time $t$, at criticality $(\tau =0)$.

In order to obtain these exponents, one can define the global persistence
probability as 
\begin{equation}
G_{M}(t)=1-\sum_{t^{\prime }=1}^{t}\rho _{m}(t^{\prime })
\end{equation}%
and 
\begin{equation}
G_{P}(t)=1-\sum_{t^{\prime }=1}^{t}\rho _{p}(t^{\prime }),
\end{equation}%
where $\rho _{m}(t^{\prime })$ and $\rho _{p}(t^{\prime })$ are the
fractions of samples that have changed the sign of their magnetization and
polarization, respectively, for the first time at the instant $t^{\prime }$.
Here, the simulations are performed for some predefined values of the
initial magnetization $m_{0}<<1$ and polarization $p_{0}<<1$. Hence, a sharp
preparation of the initial states is needed to obtain precise values for
them. After obtaining the exponents $\theta _{g_{m}}$ and $\theta _{g_{p}}$
for each value of $m_{0}$ and $p_{0}$, respectively, the final values are
achieved by performing the limit procedures $m_{0}\rightarrow 0$ and $%
p_{0}\rightarrow 0$.

In this paper, we consider the following values for $m_{0}$ and $p_{0}$:
0.002, 0.004, 0.006, and 0.008. To obtain these values, we first insert
randomly, at each site of the sublattices, a spin variable that takes the
values $\pm 1$. After that, the magnetization of the sublattices and the
polarization of the system are measured by using Eq. (\ref{orderparameters}%
). Then, spin variables are chosen randomly and their sign are changed until
we obtain a null value for the magnetizations and polarization. The last
procedure is to change the signs of $\delta /2$ sites of each sublattice, at
random, to obtain the desired initial magnetization $m_{0}$ and polarization 
$p_{0}$.

\begin{figure}[h]
\begin{center}
\includegraphics[width=\columnwidth]{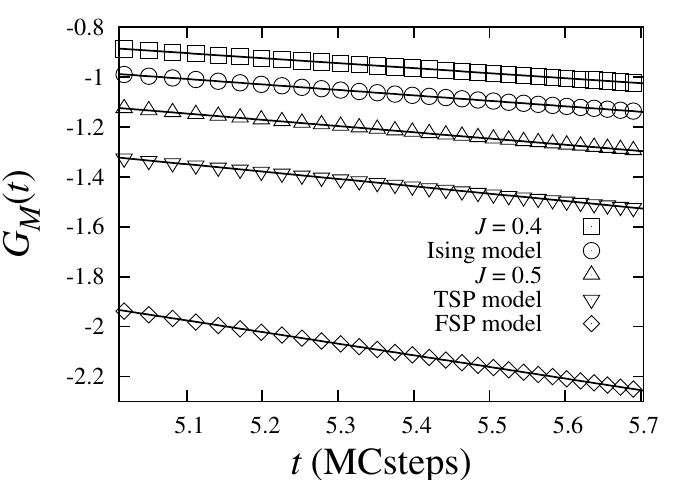} \ \includegraphics[width=%
\columnwidth]{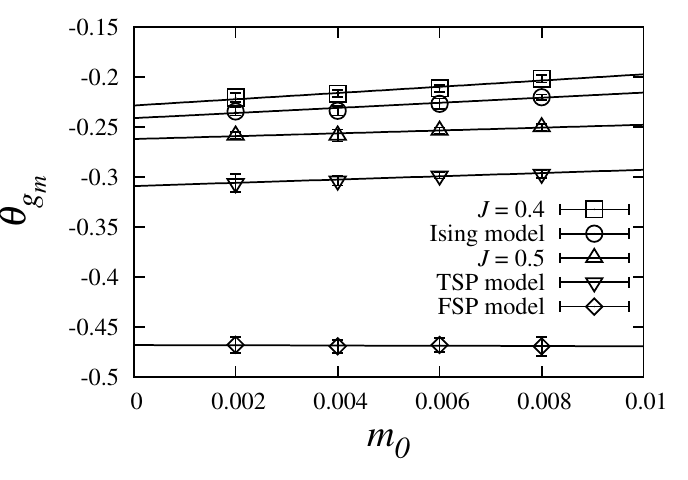}
\end{center}
\caption{Global persistence probability of the magnetization for $L=256$. On
top, polynomial decay of $G_{M}(t)\times t$ for $m_{0}=0.008$. At the
bottom, linear fit of the estimates obtained for different values of $m_{0}$%
. }
\label{fig:pgtm}
\end{figure}

In Fig. \ref{fig:pgtm} we show the decay of the global persistence
probability of the magnetization (on top) for the five considered points,
for $L=256$ and $m_{0}=0.008$. The error bars are smaller than the symbols. 
In that same figure, at the bottom, we present the plots of $\theta _{g_{m}}$
as function of $m_{0}$, as well as the limit procedure $m_{0}\rightarrow 0$.

Table \ref{table:pgtm} presents the results obtained from the limit
procedure for the three lattices, $L=64$, 128, and 256. 
\begin{table}[tbh]
\centering%
\begin{tabular}{cccc}
\hline
$J$ & $L=64$ & $L=128$ & $L=256$ \\ \hline
0.4 & 0.2063(26) & 0.2211(16) & 0.2283(28) \\ 
Ising model & 0.2186(19) & 0.2381(40) & 0.2409(26) \\ 
0.5 & 0.2417(25) & 0.2656(7) & 0.2618(14) \\ 
TSP model & 0.2835(18) & 0.3032(20) & 0.3089(23) \\ 
FSP model & 0.4678(38) & 0.4763(60) & 0.4679(13) \\ \hline
\end{tabular}%
\caption{The global persistence exponent $\protect\theta _{g_{m}}$ for the
five considered points.}
\label{table:pgtm}
\end{table}

The results show that the dynamic critical exponent $\theta _{g_{m}}$ grows
monotonically with $J$. Moreover, the values obtained for the Ising, TSP,
and FSP models can be compared with results obtained previously and found in
literature.

For the Ising critical point, the uncoupled point, our result is in complete
agreement with that presented by Schulke \textit{et al.} \cite{Schulke1997}, 
$\theta _{g}=0.238(3)$. Our result can also be compared to the value
obtained by Majumdar \textit{et al.} \cite{Majumdar1996} using a finite-size
scaling technique. By starting from a random initial configuration and
collapsing the data, they found $\theta _{g}z=0.505(20)$. If we consider our
estimate for $z_{m}$ (presented in Section \ref{Sec:z}), $z_{m}=2.156(11)$,
one finds $\theta _{g}=0.234(10)$. This result is slightly smaller than the
value obtained in this paper but it is in agreement with each other when
considering the statistical errors.

For the TSP critical point, our result should be compared to the value $%
\theta _{g}=0.350(1)$ also obtained in Ref. \cite{Schulke1997}. This
estimate is larger than ours even when one considers the statistical errors.
Then, as occurs with other models and exponents, maybe further studies are
needed in order to allow a comparison of the results.

For the FSP model, Fernandes \textit{et al.} \cite{Fernandes2006b} obtained $%
\theta _{g}=0.474(7)$ and Arashiro \textit{et al.} \cite{Arashiro2009} found 
$\theta _{g}=0.475(5)$ for the FSP model and $\theta _{g}=0.471(5)$ for the $%
n=3$ Turban model (this model belongs to the $q=4$ Potts model universality
class.) Therefore, our estimate is in good agreement with those ones
obtained previously.

\begin{figure}[h]
\begin{center}
\includegraphics[width=\columnwidth]{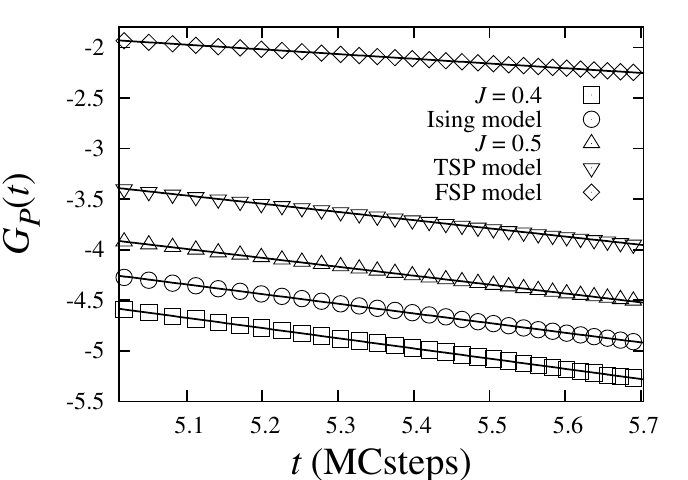} \ \includegraphics[width=%
\columnwidth]{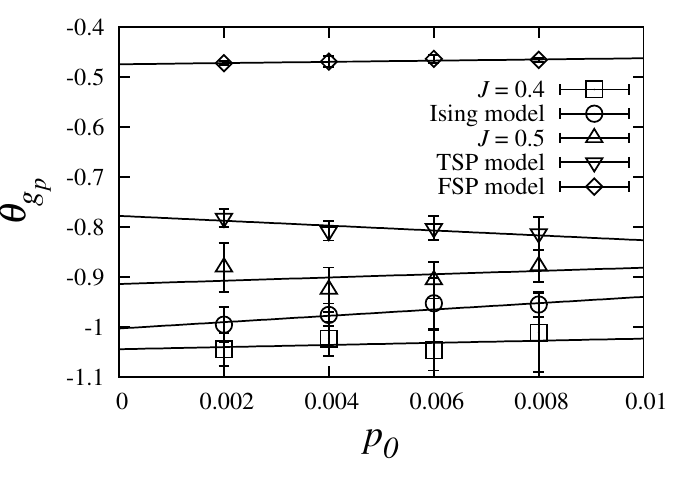}
\end{center}
\caption{Global persistence probability of the polarization for $L=256$. On
top: polynomial decay of $G_{P}(t)\times t$ for $p_{0}=0.008$. At the
bottom: linear fit of the estimates obtained for different values of $p_{0}$%
. }
\label{fig:pgtp}
\end{figure}

Fig. \ref{fig:pgtp} shows the global persistence probability in double-log
scale for the polarization, for the five points along the self-dual critical
line (on top), $L=256$ and $p_{0}=0.008$. The error bars are smaller than
the symbols. 
The plots of $\theta _{g_{p}}$ as function of $p_{0}$, as well as the limit
procedure $p_{0}\rightarrow 0$ are shown at the bottom of this figure and
the extrapolated values are presented in Table \ref{table:pgtp}. 
\begin{table}[tbh]
\centering%
\begin{tabular}{cccc}
\hline
$J$ & $L=64$ & $L=128$ & $L=256$ \\ \hline
0.4 & 1.143(55) & 1.035(12) & 1.044(19) \\ 
Ising model & 0.9941(76) & 0.9867(57) & 1.0027(80) \\ 
0.5 & 0.9279(186) & 0.9186(114) & 0.9137(346) \\ 
TSP model & 0.8183(224) & 0.7988(94) & 0.777(121) \\ 
FSP model & 0.4607(105) & 0.4570(106) & 0.4741(18) \\ \hline
\end{tabular}%
\caption{The global persistence exponent $\protect\theta _{g_{p}}$ for the
five considered points.}
\label{table:pgtp}
\end{table}

For the polarization, the global persistence exponent decreases
monotonically with $J$ showing, as above, the nonuniversal character of the
model. The values of the exponent are higher than for $\theta _{g_{m}}$ but
this difference disappears for the $q=4$ Potts critical point whereas in
this point $K=K_{4}$ and both $\theta _{g_{m}}$ and $\theta _{g_{p}}$ share
the same value.

\subsection{Dynamic critical exponents $\protect\theta _{m}$ and $\protect%
\theta _{p}$}

\label{Sec:exponents_theta}

\begin{figure}[h]
\begin{center}
\includegraphics[width=\columnwidth]{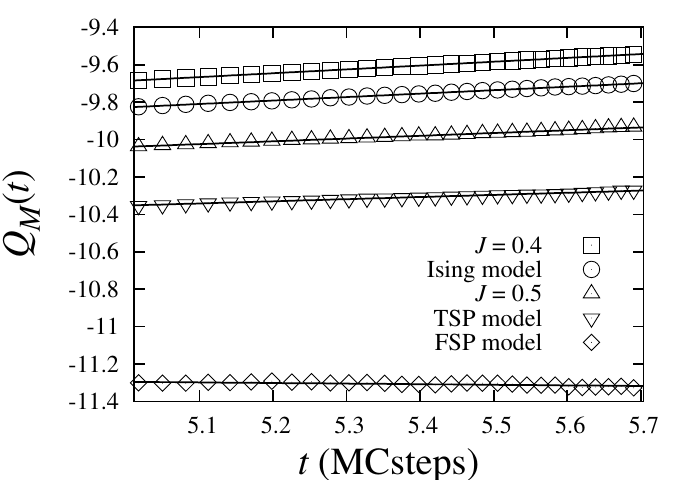}
\end{center}
\caption{The time evolution of the time correlation of the magnetization $%
Q_{M}(t)$.}
\label{fig:qtm}
\end{figure}

As stressed before, we consider two different approaches to estimate the
exponents $\theta _{m}$ and $\theta _{p}$. Our first attempt is related to
the time correlation of the magnetization and polarization, Eqs. (\ref{theta}%
) and (\ref{theta-1}), respectively. However, the huge fluctuations, even
for 300,000 samples, prevented us of considering this technique to obtain $%
\theta _{p}$. In Fig. \ref{fig:qtm}, $Q_{M}(t)$ is plotted in double-log
scale for five different values of $J$. 
Table \ref{table:qtm} shows our numerical results for each $\theta _{m}$
with the corresponding error, for the three lattice sizes considered in this
paper. 
\begin{table}[tbh]
\centering%
\begin{tabular}{cccc}
\hline
$J$ & $L=64$ & $L=128$ & $L=256$ \\ \hline
0.4 & 0.207(15) & 0.208(13) & 0.205(10) \\ 
Ising model & 0.188(20) & 0.189(17) & 0.188(13) \\ 
0.5 & 0.163(10) & 0.158(17) & 0.162(20) \\ 
TSP model & 0.129(19) & 0.131(25) & 0.121(17) \\ 
FSP model & -0.087(85) & -0.071(77) & -0.031(51) \\ \hline
\end{tabular}%
\caption{The dynamic critical exponent $\protect\theta _{m}$ obtained from
the time correlation of the magnetization, $Q_{M}(t)$, for the five coupling
constants of the Baxter model at the self-dual critical line of the
Ashkin-Teller model.}
\label{table:qtm}
\end{table}

The second method consists of calculating the exponents $\theta _{m}$ and $%
\theta _{p}$ for different values of $m_{0}<<0$ and $p_{0}<<0$,
respectively, by using the Eqs. (\ref{thtm}) and (\ref{thtp}). Their final
values are then obtained by carrying out the limit procedure $%
m_{0}\rightarrow 0$ and $p_{0}\rightarrow 0$. In order to avoid huge
fluctuations of the order parameters, which arise when $m_{0}$ and $p_{0}$
are very close to zero, we consider the following values of $m_{0}$ and $%
p_{0}$: 0.02, 0.04, 0.06, and 0.08.

\begin{figure}[h]
\begin{center}
\includegraphics[width=\columnwidth]{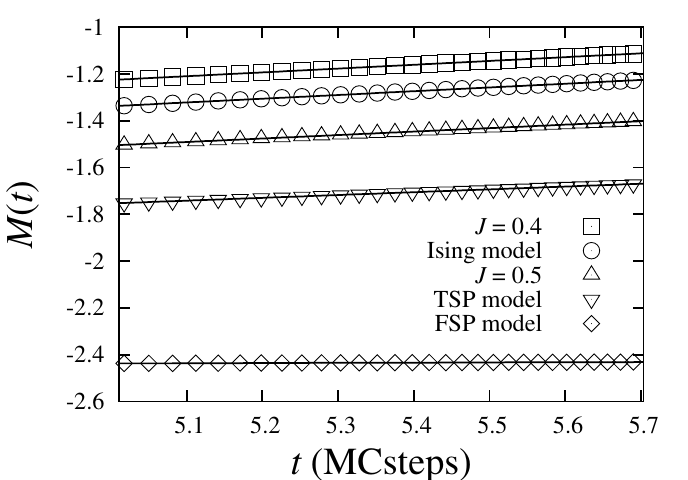} \ \includegraphics[width=%
\columnwidth]{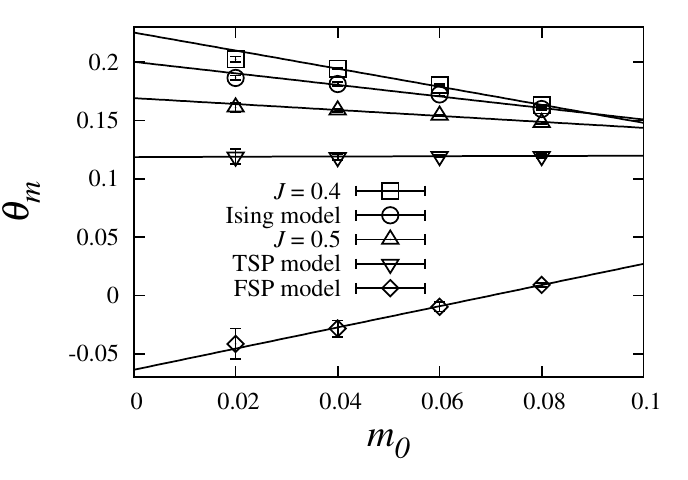}
\end{center}
\caption{The plot of $M(t)\times t$, Eq. (\protect\ref{thtm}), for the
initial magnetization $m_{0}=0.08$ and $L=256$. On top: polynomial behavior
of $M(t)\times t$. At the bottom: linear fit of the estimates obtained for
different values of $m_{0}$.}
\label{fig:thtm}
\end{figure}

Fig. \ref{fig:thtm} shows the polynomial behavior of $M(t)\times t$ for $%
m_{0}=0.06$ in double-log scales, for the five different values of $J$ and $%
L=256$ (on top).

The limit procedures are shown at the bottom of this figure and the
extrapolated values can be seen in Table \ref{table:thtm} for the three
lattices, $L=64$, $128$, and $256$. 
\begin{table}[tbh]
\centering%
\begin{tabular}{cccc}
\hline
$J$ & $L=64$ & $L=128$ & $L=256$ \\ \hline
0.4 & 0.2171(23) & 0.2198(35) & 0.2203(45) \\ 
Ising model & 0.1915(16) & 0.1999(20) & 0.1951(26) \\ 
0.5 & 0.1681(64) & 0.1646(33) & 0.1666(14) \\ 
TSP model & 0.1172(19) & 0.1196(9) & 0.1176(10) \\ 
FSP model & -0.0580(21) & -0.0707(59) & -0.0611(42) \\ \hline
\end{tabular}%
\caption{The dynamic critical exponent $\protect\theta _{m}$ for the five
considered points.}
\label{table:thtm}
\end{table}

Our results displayed in Tables \ref{table:qtm} and \ref{table:thtm} are in
good agreement with each other and show that the exponent $\theta _{m}$
varies continuously with $J$. The estimates also corroborate the available
values for the Ising and FSP models. For the former one, our results should
be compared with those ones showed by Grassberger \cite{Grassberger1995}, $%
\theta =0.191(3)$, Li \textit{et al.} \cite{Li1997}, $\theta =0.191(2)$, and
Okano \textit{et al.} \cite{Okano1997}, $\theta =0.191(1)$. For the FSP
model, Okano \textit{et al.} \cite{Okano1997} conjectured that the exponent $%
\theta $ should be negative and close to zero and the results for this model 
\cite{Silva2004,Arashiro2009} as well as for the Ising model with three-spin
interactions \cite{Arashiro2009} validate this assertion. Besides the $q=4$
Potts model, there have been shown in some papers that there are models in
which the exponent $\theta $ can also have a negative value, for instance,
the tricritical Ising model \cite{Janssen1994}, Blume-Capel model \cite%
{Silva20022}, metamagnetic model \cite{Silva2013}, and Baxter-Wu model \cite%
{Arashiro2003,Hadjiagapiou2005}. On the other hand, our estimate for the TSP
model is completely different from some values published up to now,
0.0815(27) \cite{Schulke1995} and 0.075(3) \cite{Schulke1997a}.
Nevertheless, our estimate, is in agreement with the result for the same
point in the critical self-dual line of the Ashkin-Teller model \cite{Li1997}%
. In that work, the authors do not estimate directly the exponent $\theta
_{m}$ for the critical point of the TSP model ($y=3/4$). However, as pointed
out by Chatelain \cite{Chatelain2004}, this exponent varies roughly linearly
with the parameter $y$ which allows us to estimate the exponent $\theta _{m}$
in this point, leading to $\theta _{m}\approx 0.111$. This result is
compatible with ours for the critical point of the TSP model.

In order to obtain the exponent $\theta _{p}$, we first consider the same
initial conditions, i.e., $p_{0}=0.02$, $0.04$, $0.06$ and $0.08$. Fig. \ref%
{fig:thtp} displays the behavior of $P(t)\times t$, Eq. (\ref{thtp}), in
double-log scale for $p_{0}=0.08$ and $L=256$ for the five critical points
considered. 
\begin{figure}[h]
\begin{center}
\includegraphics[width=\columnwidth]{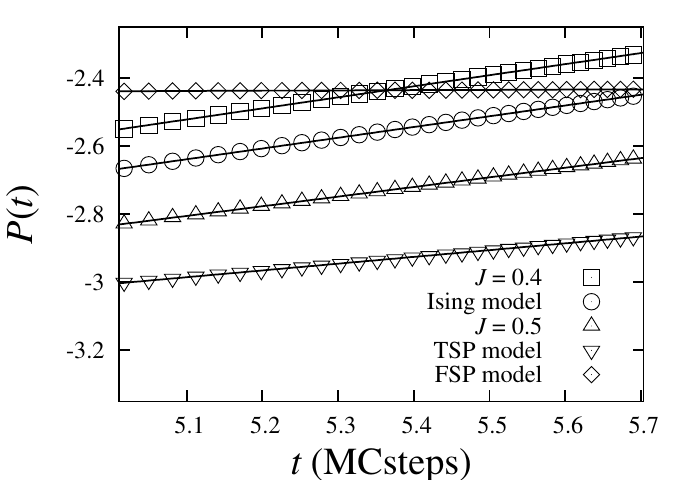} \ \ \includegraphics[width=%
\columnwidth]{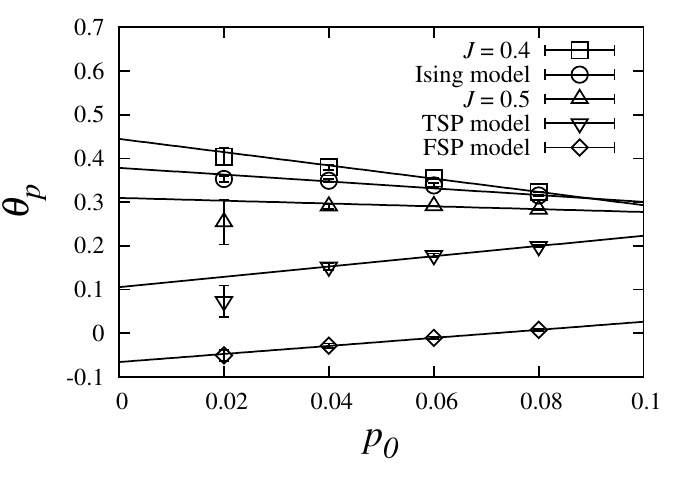}
\end{center}
\caption{The plot of $P(t)\times t$, Eq. (\protect\ref{thtp}), for the
initial polarization $p_{0}=0.08$ and $L=256$. On top: polynomial behavior
of $P(t)\times t$. At the bottom: linear fitting of the estimates obtained
for different values of $p_{0}$.}
\label{fig:thtp}
\end{figure}
The extrapolated values, obtained from the limit procedure $p_{0}\rightarrow
0$, are presented in Table \ref{table:thtp} for the three lattices, $L=64$,
128, and 256. 
\begin{table}[tbh]
\centering%
\begin{tabular}{cccc}
\hline
$J$ & $L=64$ & $L=128$ & $L=256$ \\ \hline
0.4 & 0.4466(320) & 0.4364(16) & 0.4301(7) \\ 
Ising model & 0.4317(39) & 0.3857(94) & 0.3638(58) \\ 
0.5 & 0.2797(361) & 0.3017(125) & 0.2860(115) \\ 
TSP model & 0.1403(372) & 0.0937(210) & 0.1106(59) \\ 
FSP model & -0.0611(115) & -0.0597(44) & -0.0645(1) \\ \hline
\end{tabular}%
\caption{The dynamic critical exponent $\protect\theta _{p}$ for the five
considered points.}
\label{table:thtp}
\end{table}

The results show that the exponent $\theta_p$ decreases monotonically with
respect to $J$. They are completely different from those obtained by Li 
\textit{et al.} \cite{Li1997}, but for the $q=4$ Potts critical point ($y=1$
in that paper). They showed that the polarization is negative for all
considered points.

\subsection{The dynamic critical exponents $z_m$ and $z_p$}

\label{Sec:z}

Finally, the dynamic critical exponents $z_{m}$ and $z_{p}$ are obtained by
combining results from samples submitted to different initial conditions
(ordered state for the order parameter and disordered one for the second
moment of the order parameter), Eqs. (\ref{f2m}) and (\ref{f2p}), where the
dimension of the system is $d=2$. This technique has proven to be very
efficient in estimating the exponent $z$ for a large number of models \cite%
{Silva20021,Silva20022,Arashiro2003,Fernandes2005,Fernandes2006b,Arashiro2009}%
.

\begin{figure}[h]
\begin{center}
\includegraphics[width=\columnwidth]{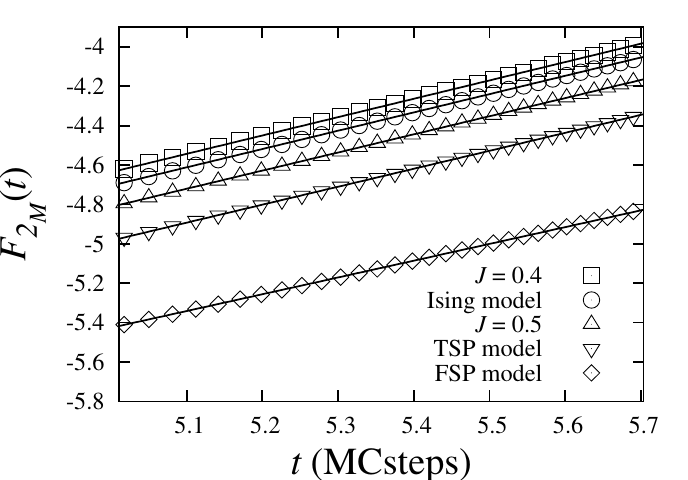}
\end{center}
\caption{Time evolution of $F_{2_{M}}(t)$ for the five coupling constants
and $L=256$.}
\label{fig:f2tm}
\end{figure}

The time evolution of $F_{2_{M}}$, obtained from Eq. (\ref{f2m}), is shown
in Fig. (\ref{fig:f2tm}) in double-log scale for the five considered points
and $L=256$. The error bars are smaller than the symbols. 

The mean values of $z_{m}$ and the corresponding errors are given in Table %
\ref{table:f2tm} for $L=64$, 128, and 256. 
\begin{table}[tbh]
\centering%
\begin{tabular}{cccc}
\hline
$J$ & $L=64$ & $L=128$ & $L=256$ \\ \hline
0.4 & 2.113(18) & 2.139(16) & 2.147(12) \\ 
Ising model & 2.129(12) & 2.156(11) & 2.156(11) \\ 
0.5 & 2.154(9) & 2.168(13) & 2.172(13) \\ 
TSP model & 2.175(5) & 2.198(14) & 2.194(17) \\ 
FSP model & 2.318(11) & 2.342(21) & 2.346(21) \\ \hline
\end{tabular}%
\caption{The dynamic critical exponent $z_{m}$ for the five considered
points.}
\label{table:f2tm}
\end{table}

In the uncoupling point, $J=K=0.5\mbox{ln}(1+\sqrt{2})$, the exponent $z_{m}$
is in complete agreement with those obtained for the two-dimensional Ising
model \cite{Li1996,Okano1997}. For the TSP model, our estimate is also in
complete agreement with those ones presently accepted for the model, $%
z=2.1983(81)$ \cite{Schulke1995}, obtained from the time evolution of the
self-correlation and $z=2.197(3)$ obtained by mixing moments of the
magnetization under different initial conditions \cite{Silva20021}, $%
F_{2}(t) $. However, our estimate of $z_{m}$ for the FSP model, is larger,
but very close to the values recently obtained for that model, $z=2.290(3)$ 
\cite{Silva20021} and $z=2.294(3)$ \cite{Fernandes2006b}, for the Baxter-Wu
model \cite{Arashiro2003}, $z=2.294(6)$, and for the $n=3$ Turban model \cite%
{Arashiro2009}, $z=2.292(4)$, both belonging to the same universality class
of the FSP model.

\begin{figure}[h]
\begin{center}
\includegraphics[width=\columnwidth]{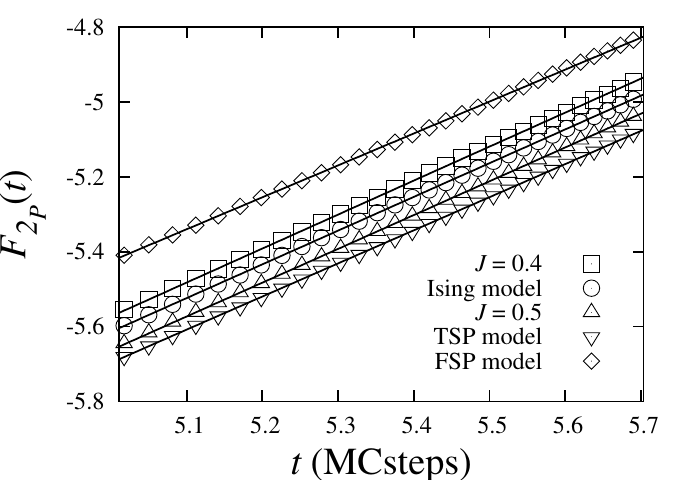}
\end{center}
\caption{Time evolution of $F_{2_{P}}(t)$ for the five coupling constants.}
\label{fig:f2tp}
\end{figure}

In Fig. \ref{fig:f2tp} we show the time dependence of $F_{2_{P}}(t)$ in
double-log scales for the five considered points and $L=256$. The error
bars, obtained from five independent runs, are smaller then the symbols. 

The linear fits of these curves, as well as of those ones with $L=64$ and $%
L=128$, lead to the values presented in Table \ref{table:f2tp}. 
\begin{table}[!htb]
\centering
\begin{tabular}{cccc}
\hline
$J$ & $L=64$ & $L=128$ & $L=256$ \\ \hline
0.4 & 1.889(19) & 2.236(16) & 2.201(8) \\ 
Ising model & 1.932(25) & 2.225(21) & 2.220(16) \\ 
0.5 & 2.004(25) & 2.232(17) & 2.212(26) \\ 
$q=3$ Potts model & 2.092(10) & 2.253(19) & 2.258(17) \\ 
$q=4$ Potts model & 2.210(12) & 2.341(12) & 2.338(38) \\ \hline
\end{tabular}%
\caption{The dynamic critical exponent $z_p$ for the five considered points.}
\label{table:f2tp}
\end{table}

By taking into account the statistical errors, the results shown in Tables %
\ref{table:f2tm} and \ref{table:f2tp} ensure that the exponents $z_{m}$ and $%
z_{p}$ are varying with respect to $J$. In this case, only the exponent for
FSP model is different from the others. Besides, for this critical point,
the exponents $z_{m}$ and $z_{p}$ share the same values (within the error
bars). On the contrary, the exponent $z_{p}$ is greater than $z_{m}$ for the
other points, in contrast with the results shown in Ref. \cite{Li1997} for
the Ashkin-Teller model, where there is no distinction between the two
critical indexes.

\subsection{Static critical exponents}

\label{Sec:static_exponents}

Here we finally calculate the static critical exponents of de AT model. By
using the exponents $z_{m}$ and $z_{p}$ obtained in the previous section, we
calculate the exponent $\nu $ for the magnetization and polarization
respectively. Differently from what occurs with the dynamic exponents, the
computing of static exponents deserve a more detailed analysis of
uncertainties and of final estimates. In this analysis we consider both the
external and internal averages.

In this paper, the static exponents were calculated$\ $by using $%
N_{run}=4000 $\ runs in order to compute the averaged time series in the
situation which the system starts from $m_{0}=1$ (or $p_{0}=1$). First, the
error bars are obtained with $N_{b}=5$\ different bins (for polarization).
For the case of magnetization we have $N_{b}=10$\ different bins since the
lattices are doubled. Here it is important to differentiate
\textquotedblleft bin\textquotedblright\ of \textquotedblleft
seed\textquotedblright . We always used 5 seeds, but due to duplicity of
lattices in the AT model, and considering the isotropic case ($K_{1}=K_{2}=K$%
) the number of bins is equal to the double of seeds for the magnetization
case.

We numerically compute the derivative through Eqs. (\ref{1niz}) and (\ref%
{1nizp}), which leads to: 
\begin{equation}
D(t)=\frac{1}{2\delta }\ln \frac{\overline{O}(K_{4}(T_{c}+\varepsilon
),K(T_{c}+\varepsilon ),t)}{\overline{O}(K_{4}(T_{c}-\varepsilon
),K(T_{c}-\varepsilon ),t)}\text{,}  \label{Eq:Derivative_expoent}
\end{equation}%
where $\overline{O}(K_{4}(T),K(T),t)$ denotes the averaged
magnetization/polarization calculated in values above and below the critical
temperatures. If $K(T_{c})=K_{c}=\frac{1}{T_{c}}$, it is interesting to
observe that a perturbation on the critical temperature, $T_{c}\pm
\varepsilon $, produces $\ K(T_{c}\pm \varepsilon )=1/(T_{c}\pm \varepsilon
)=\frac{1/T_{c}}{(1\pm \varepsilon /T_{c})}=K_{c}/(1\pm \delta )$, where $%
\delta =\varepsilon /T_{c}$. This means that when we divide $K_{c}$ by $%
(1\pm \delta )$, the critical temperature is perturbed by a value $\pm
\varepsilon =\pm \delta T_{c}$. Similarly, $K_{4}(T_{c}\pm \varepsilon
)=K_{4}^{c}/(1\pm \delta )$.

\begin{figure}[h]
\begin{center}
\includegraphics[width=\columnwidth]{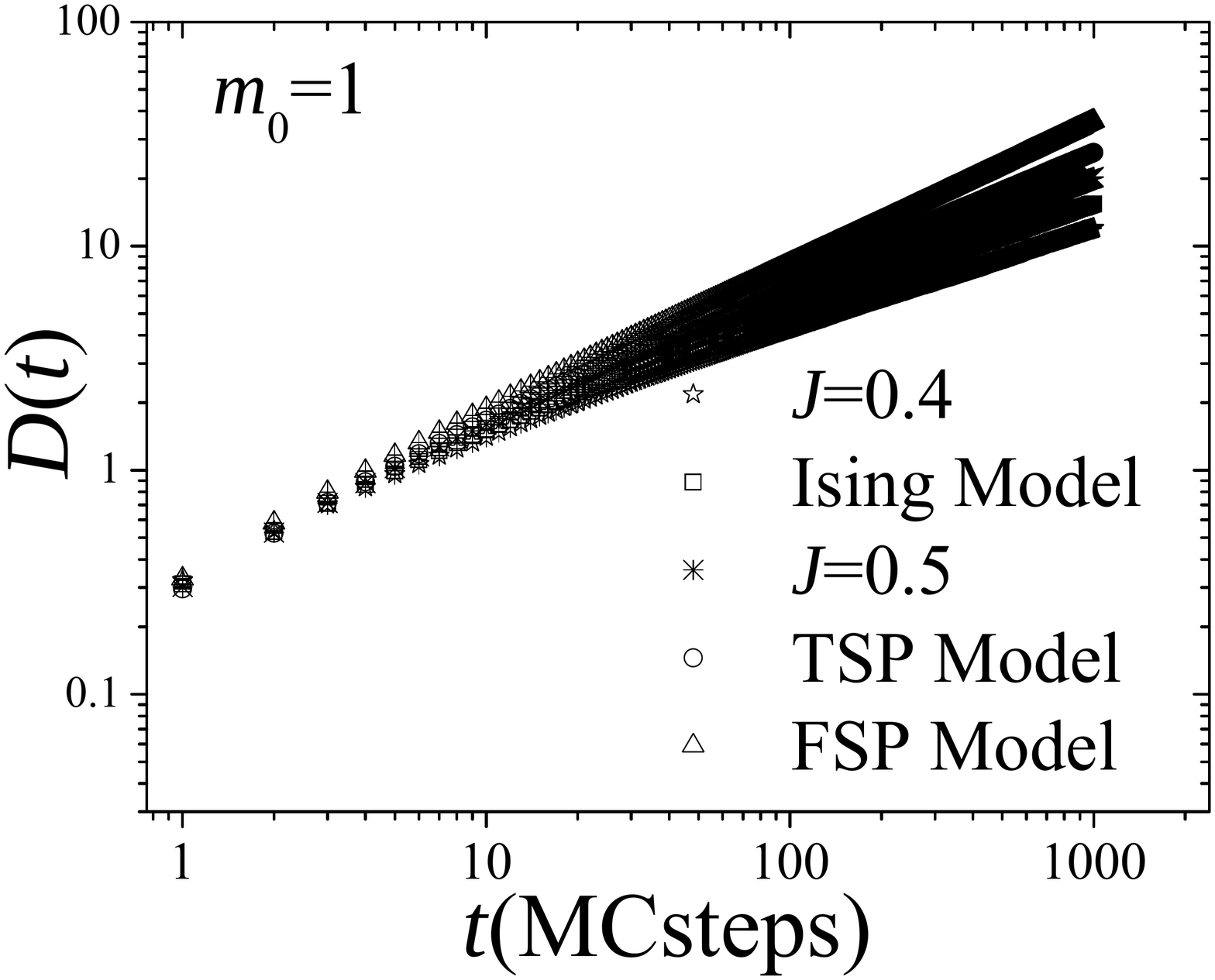} \ \includegraphics[width=%
\columnwidth]{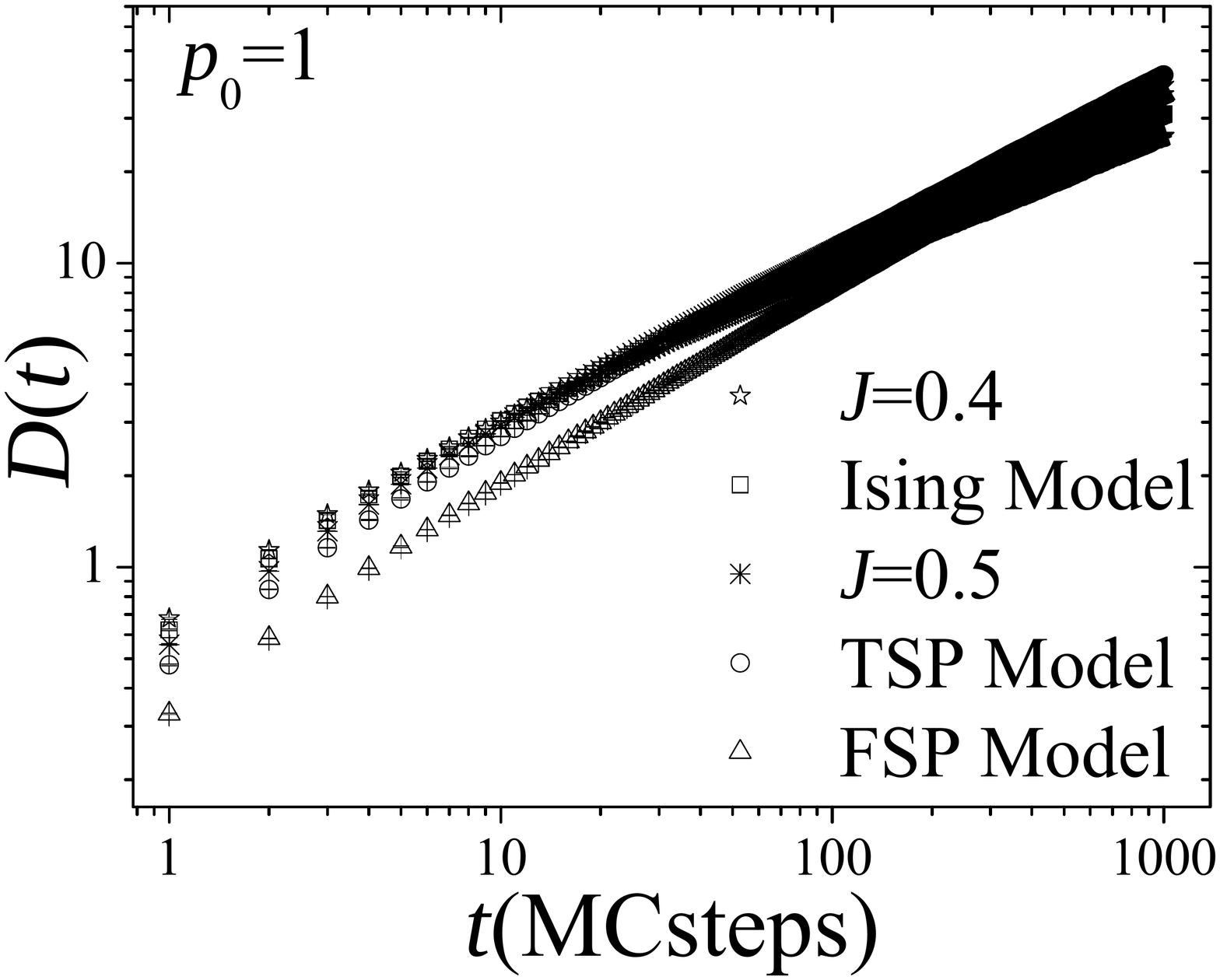}
\end{center}
\caption{Time evolution of $D(t)$ for the five coupling constants. Top plot:
magnetization. Bottom plot: Polarization.}
\label{Fig:Derivative}
\end{figure}

Fig. \ref{Fig:Derivative} shows the time evolution of $D(t)$ for
magnetization and polarization, calculated by Eq. (\ref%
{Eq:Derivative_expoent}). For the magnetization, the error bars are obtained
by using an average over different $N_{b}^{2}=100$ points while for the
polarization $N_{b}^{2}=25$ points since we cross the $N_{b}$ time series
simulated above critical parameter: $\overline{O}(T_{c}+\varepsilon ,t)$, on
the perpendicular line as previously described, with $N_{b}$ time series
simulated below critical parameter $\overline{O}(T_{c}-\varepsilon ,t)$. We
can clearly observe a power law behavior (log-log plot) for all points
studied.

So, in order to compute the static exponents, we observe that the exponent $%
\phi =1/(\nu z)$ has an important variation on the different time lags
considered, and therefore, such a variation must be considered in the final
estimates of the exponent $\nu $. In this analysis we index by $k$ the time
lag $[t_{i}^{(k)},t_{f}^{(k)}]$, where $k=1,...,n$. It was built considering
that the minimum size of the interval is $\Delta =100$ MCsteps. Moreover,
the minimum $t_{i}$ adopted is 50 MCsteps, while the maximum $t_{f}$ is 1000
MCsteps.

We prepared an algorithm that considers the same number of points per
interval which allows to perform linear fits under the same conditions for
all different intervals considered in the analysis. The apropriate number of
points per interval in this paper was $n_{p}=25$, and the spacing was
adjusted to satisfy such restriction.

Let us denote here: $\overline{O}_{l}(T_{c}+\varepsilon ,t)$ the order
parameter averaged over $N_{run}$ different runs corresponding to the $l-$th
bin calculated in $T_{c}+\varepsilon $ and $\overline{O}_{m}(T_{c}-%
\varepsilon ,t)$ corresponding to the $m-$th bin that was calculated in $%
T_{c}-\varepsilon $. Denoting $\phi _{k}^{(l,m)}$ the exponent $1/\nu z$,
calculated using the two bins $l$ and $m$ previously reported for the time
lag $k$ after a fitting of the power law $\frac{1}{2\delta }\ln \frac{%
\overline{O}_{l}(T_{c}+\varepsilon ,t)}{\overline{O}_{m}(T_{c}-\varepsilon
,t)}\propto t^{\phi _{k}^{(l,m)}}$, our final estimate of the exponent $\phi
_{k}\ $in this time lag is:%
\begin{equation}
\phi _{k}=\frac{1}{N_{b}^{2}}\sum_{l,m=1}^{N_{b}}\phi _{k}^{(l,m)}
\label{Eq:over_the_bins}
\end{equation}%
which is an average over the bins. In this case we have an uncertainty given
by 
\begin{equation}
\sigma _{k}^{2}=\frac{1}{N_{b}^{2}(N_{b}^{2}-1)}\sum_{l,m=1}^{N_{b}}(\phi
_{k}^{(l,m)}-\phi _{k})^{2}\text{.}  \label{Eq:Internal}
\end{equation}

Finally, we have the final estimate $\phi =\frac{1}{n}\sum_{k=1}^{n}\phi
_{k} $ which is an average over the time lags and which leads to a final
incertainty:%
\begin{equation}
\begin{array}{lll}
\sigma _{\phi }^{2} & = & \frac{1}{n(n-1)}\sum_{k=1}^{n}(\phi _{k}-\phi
)^{2}+\frac{1}{n^{2}}\sum_{k=1}^{n}\sigma _{k}^{2} \\ 
&  &  \\ 
& = & \sigma _{ext}^{2}+\sigma _{int}^{2}%
\end{array}
\label{Eq:nu_final_uncertainty}
\end{equation}%
where the first term of the right side corresponds to external uncertainty
(variation over the time lags). This term is a kind of geographic variation
of the exponent. The second part corresponds to the internal component of
uncertainty, i.e., the variation over the pairs of the different seeds for
each time lag. For the more skeptical, we also elaborate a bootstrapping
version of this analysis. In this case we choose two sets of 5 seeds (which
should be repeated as prescribed by the bootstrapping method) and compose
two new time series by averaging them. The first one corresponding to the
pararameter $T_{c}+\varepsilon $ and the other one corresponding to the
parameter $T_{c}-\varepsilon $ yields an exponent. We can repeat this
procedure $N_{sample}$ times instead of taking the $N_{b}^{2}$ possible
pairs. So we can replace the Eqs. (\ref{Eq:over_the_bins}) and (\ref%
{Eq:Internal}) by:$\ \phi _{k}=\frac{1}{Nsample}\sum_{i_{s}=1}^{Nsample}\phi
_{k}^{(i_{s})}$ and $\sigma _{k}^{2}=\frac{1}{(N_{sample}-1)}%
\sum_{i_{s}=1}^{N_{sample}}(\phi _{k}^{(i_{s})}-\phi _{k})^{2}$
respectively, where $\phi _{k}^{(i_{s})}$ denotes the exponent calculated
for the $i_{s}$-th element of the sample for the $k-$th time lag. The Eq. (%
\ref{Eq:nu_final_uncertainty}) remains the same. Given the exponent $z\pm
\sigma _{z}$ previously calculated, the final estimate of $\nu $ is obtained
as $\nu =(z\phi )^{-1}$, and the uncertainty is obtained by $\sigma _{\nu
}=\nu \sqrt{\left( \frac{\sigma _{z}}{z}\right) ^{2}+\left( \frac{\sigma
_{\phi }}{\phi }\right) ^{2}}$.

\begin{center}
\begin{figure*}[h]
\begin{center}
\includegraphics[width=\columnwidth]{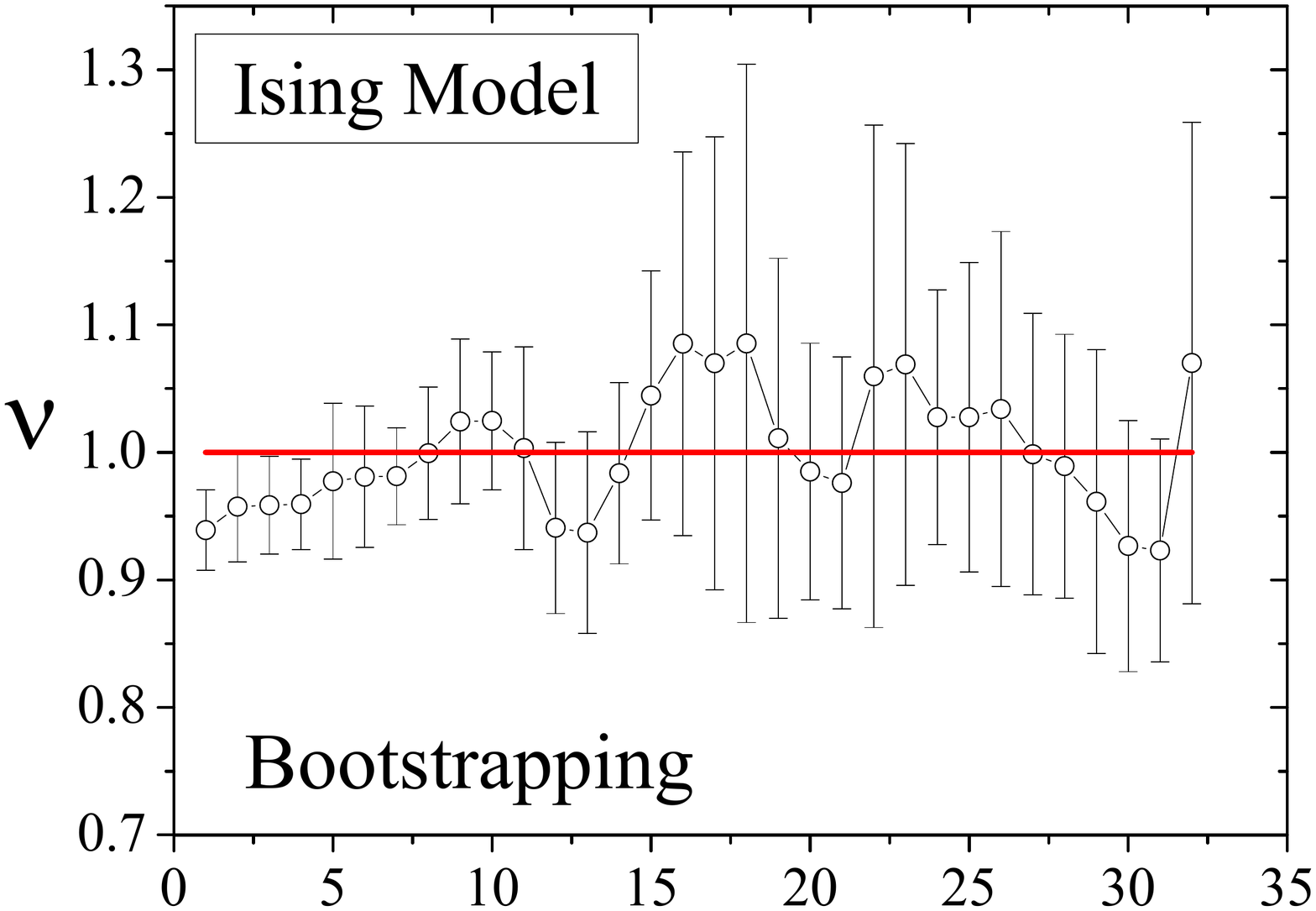}%
\includegraphics[width=\columnwidth]{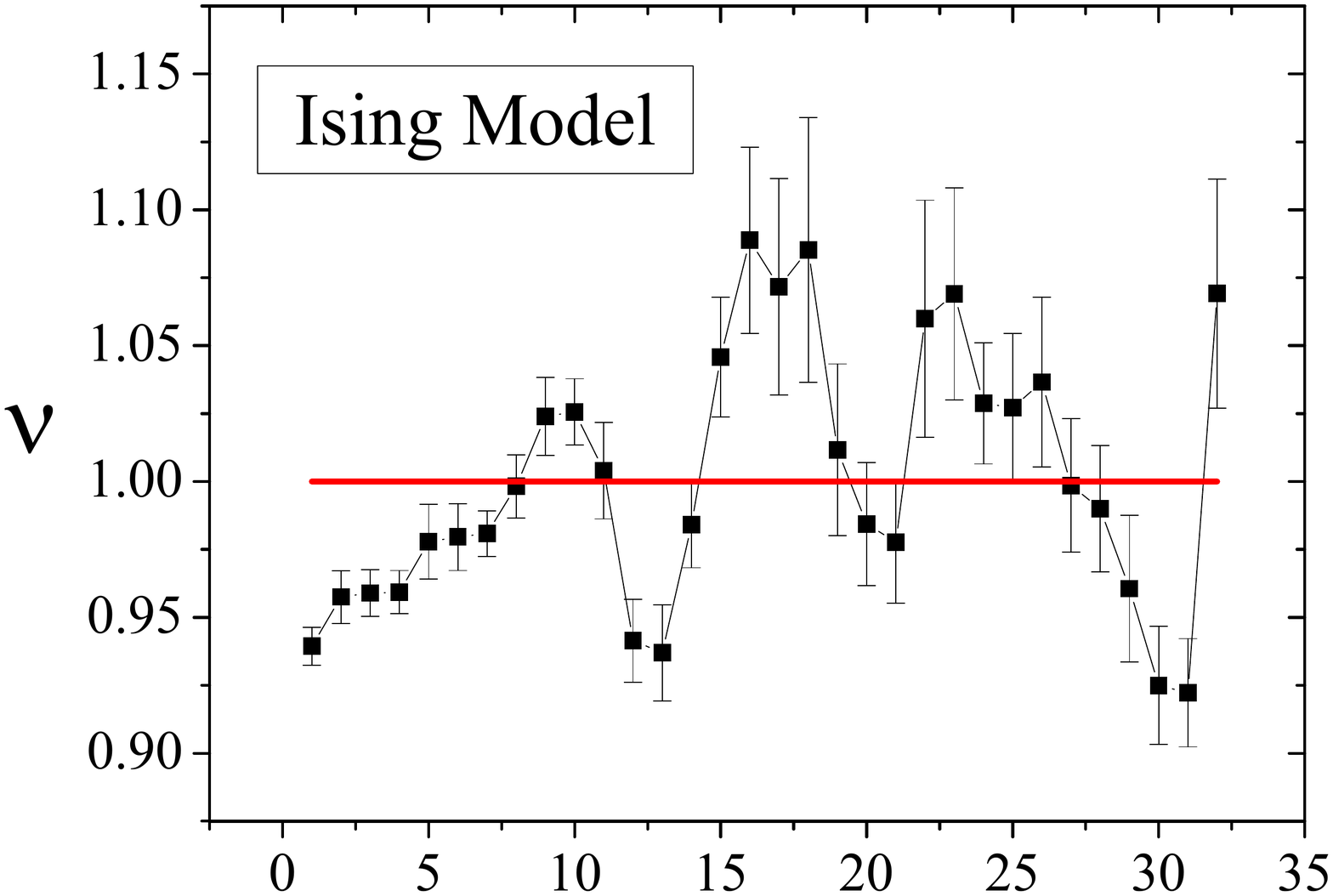} %
\includegraphics[width=\columnwidth]{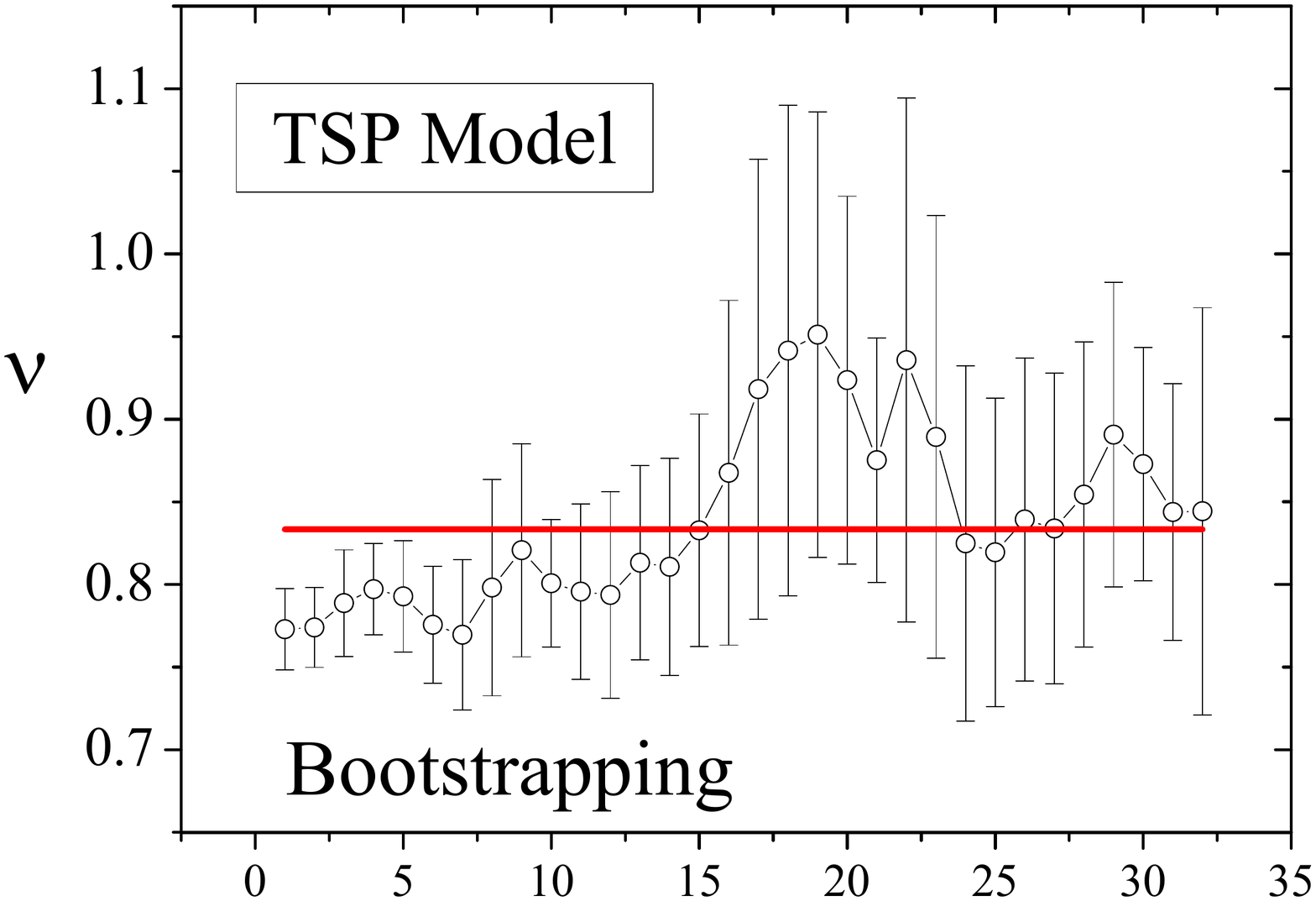}%
\includegraphics[width=\columnwidth]{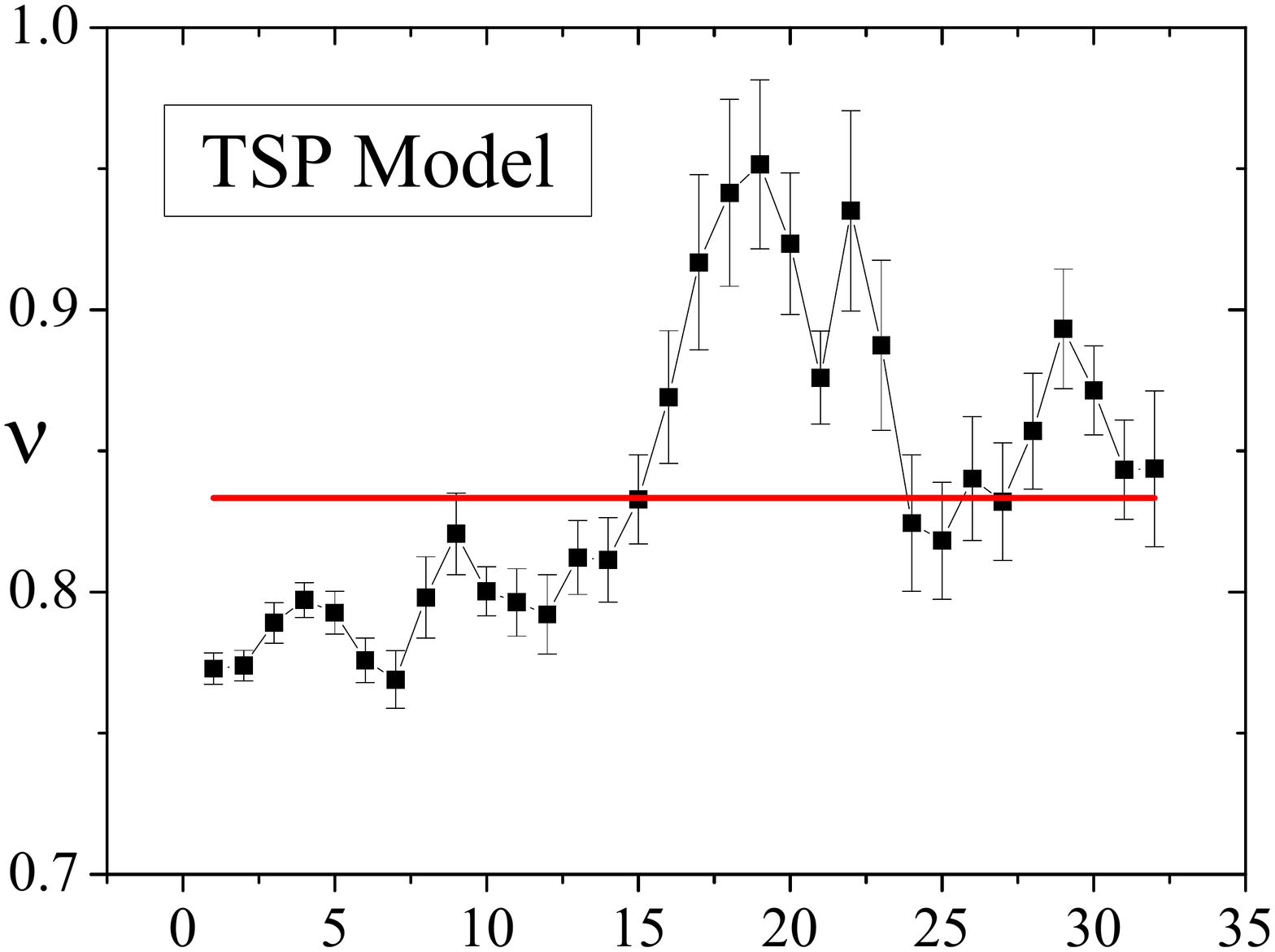} %
\includegraphics[width=\columnwidth]{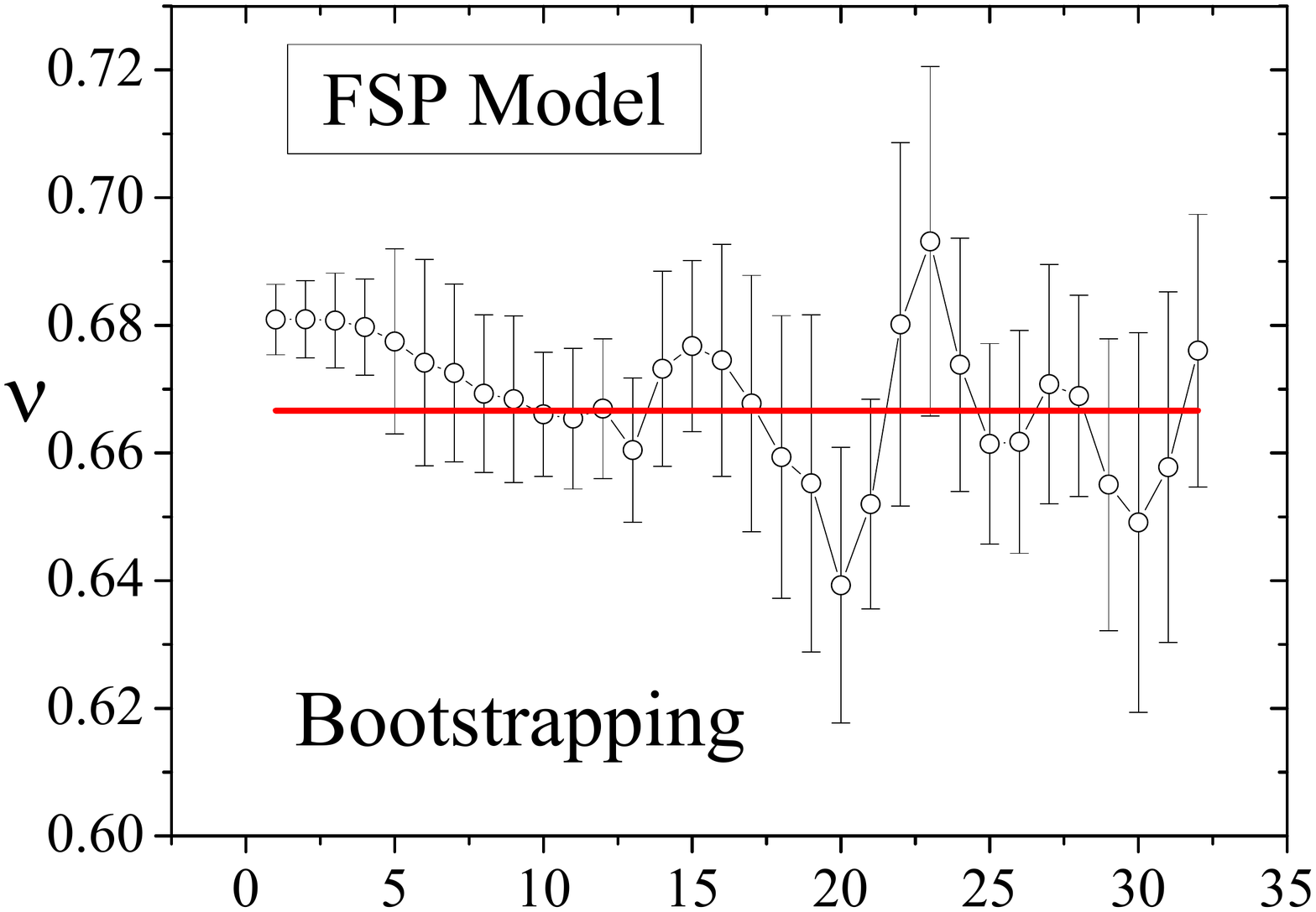}%
\includegraphics[width=\columnwidth]{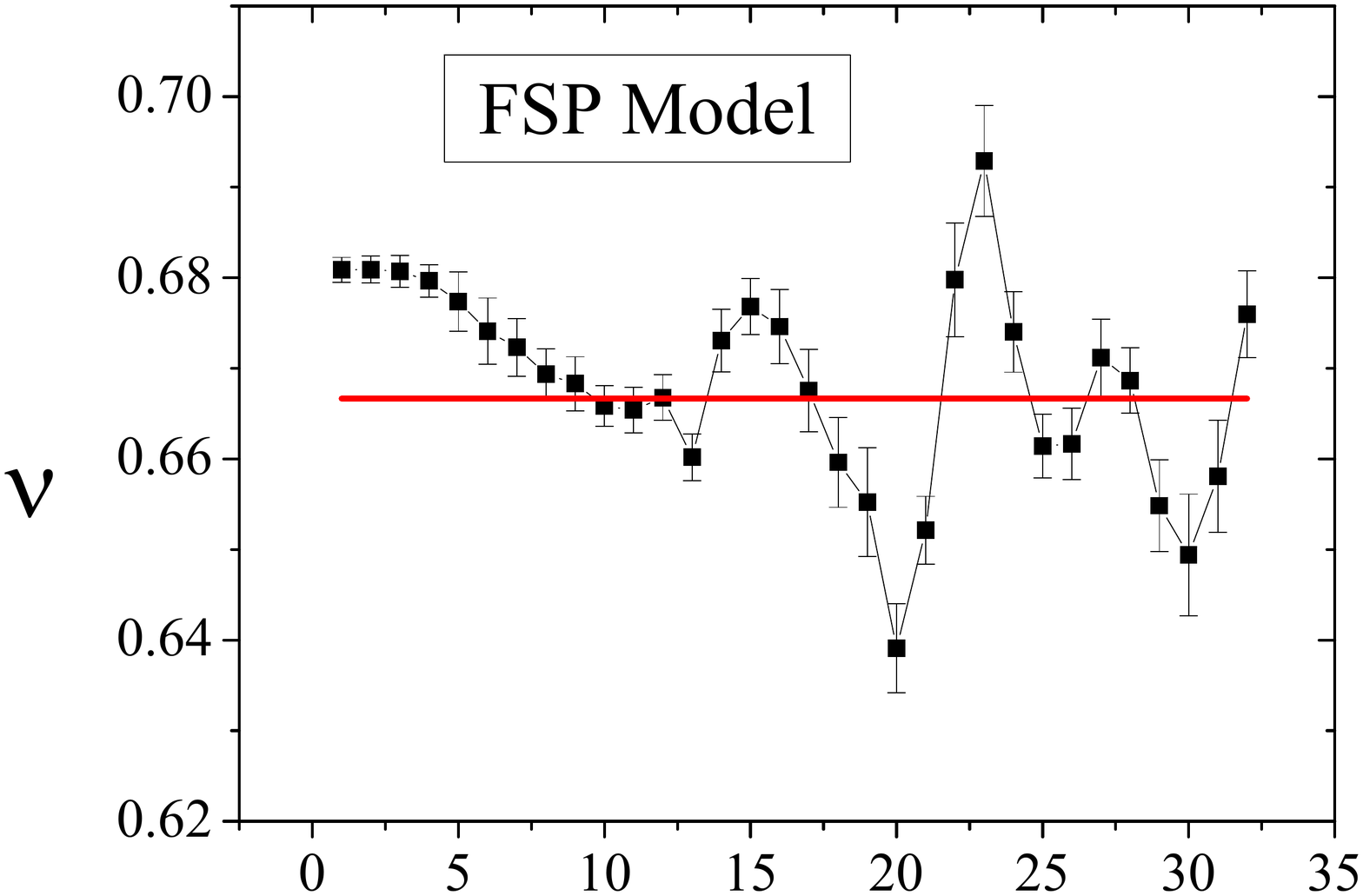}
\end{center}
\caption{Estimates of the exponents $\protect\nu _{m}$ for the different
time lags. The left-side plots show the exponents obtained by bootstrapping
method by adopting $N_{sample}=10^{3}$ while the right-side ones were
obtained with a simple crossing of $N_{b}=10\ $seeds (100 points). }
\label{Fig:exponents_nu}
\end{figure*}
\end{center}

First of all, in order to observe the variation of the exponent $\nu $ over
the different time lags we prepare a plot to show the variation of $\nu $
and its respective error bars for the different parts (time lags) of the
power law (Fig. \ref{Fig:exponents_nu}). The $x-$axis denotes a number that
indexes one specific time lag. It is important to mention that, it is a
simple ordering whereas we do not know which time lag corresponds to the
specific exponent since we are interested only in observing the fluctuations
of this exponent. In this figure, we show the exponents of the magnetization
for the points corresponding to: Ising model, TSP, and FSP models where the
right side corresponds to estimates obtained by using bootstrapping and the
left side the regular method (both previously described).

This plot shows that the estimates can be deeply changed along the power law
but the theoretical prediction is corroborated. The bootstrapping method
produces higher error bars as expected. So, taking into account the
different source variations, we obtain estimates to the exponent $\nu $ for
the different points studied in this paper. In Table \ref{Table:exponents_nu}
we present our final estimates of this exponent. The term (boot) refers to
exponents obtained using bootstraping. The terms \textit{max} and \textit{min%
} mean the largest and smallest values found in our analysis. The
conjectured values are shown in the las column and denoted by an asterisk,
and are expected to share the same value for both magnetization ($m$) and
polarization ($p$). We can observe a good agreement between the conjectured
values and our estimates.

\begin{table*}[tbh]
\centering%
\begin{tabular}{lllllllllll}
\hline
$J$ & $\nu _{m}$ & $\nu _{m}^{(boot)}$ & $\nu _{m}^{(\min )}$ & $\nu
_{m}^{(\max )}$ & $\nu _{p}$ & $\nu _{p}^{(boot)}$ & $\nu _{p}^{(\min )}$ & $%
\nu _{p}^{(\max )}$ & $\nu ^{\ast }$ & $\delta _{best}$ \\ \hline
0.4 & 1.060(12) & 1.060(24) & 1.009(28) & 1.117(30) & 1.021(12) & 1.021(16)
& 0.997(48) & 1.066(52) & $1.08900$ & $0.005$ \\ 
Ising model & 1.001(19) & 1.000(42) & 0.922(38) & 1.089(68) & 0.974(12) & 
0.974(28) & 0.911(14) & 1.020(88) & $1.00000$ & $0.002$ \\ 
0.5 & 0.893(14) & 0.893(24) & 0.844(19) & 0.941(46) & 0.862(11) & 0.862(18)
& 0.820(20) & 0.912(42) & $0.91500$ & $0.002$ \\ 
TSP model & 0.839(19) & 0.839(36) & 0.773(11) & 0.951(58) & 0.807(20) & 
0.807(28) & 0.749(15) & 0.904(86) & $0.8333\overline{3}$ & $0.001$ \\ 
FSP model & 0.6682(41) & 0.6684(74) & 0.6391(98) & 0.693(12) & 0.6679(58) & 
0.6680(82) & 0.637(11) & 0.687(17) & $0.6666\overline{6}$ & $0.002$ \\ \hline
\end{tabular}%
\caption{The static critical exponents $\protect\nu $ for the five
considered points, for magnetization ($m$) and polarization ($p$).\ All
estimates were obtained for the largest lattice used in this work: $L=256$. }
\label{Table:exponents_nu}
\end{table*}

It is important to mention that it is the first time that such exponents
have been obtained by MC simulations and to the best of our knowledge, even
for equilibrium MC simulations. The agreement between the exponents $\nu \ $%
for the polarization and magnetization was only a conjecture.

By following the same process, we analyze the decay of magnetization and
polarization described by Eqs. (\ref{Eq:Decay_mag}) and (\ref{Eq:Decay_pol}%
). The time evolving of these amounts are shown in Fig. \ref%
{Fig:time_decay_mag_pol}.

\begin{figure}[h]
\begin{center}
\includegraphics[width=\columnwidth]{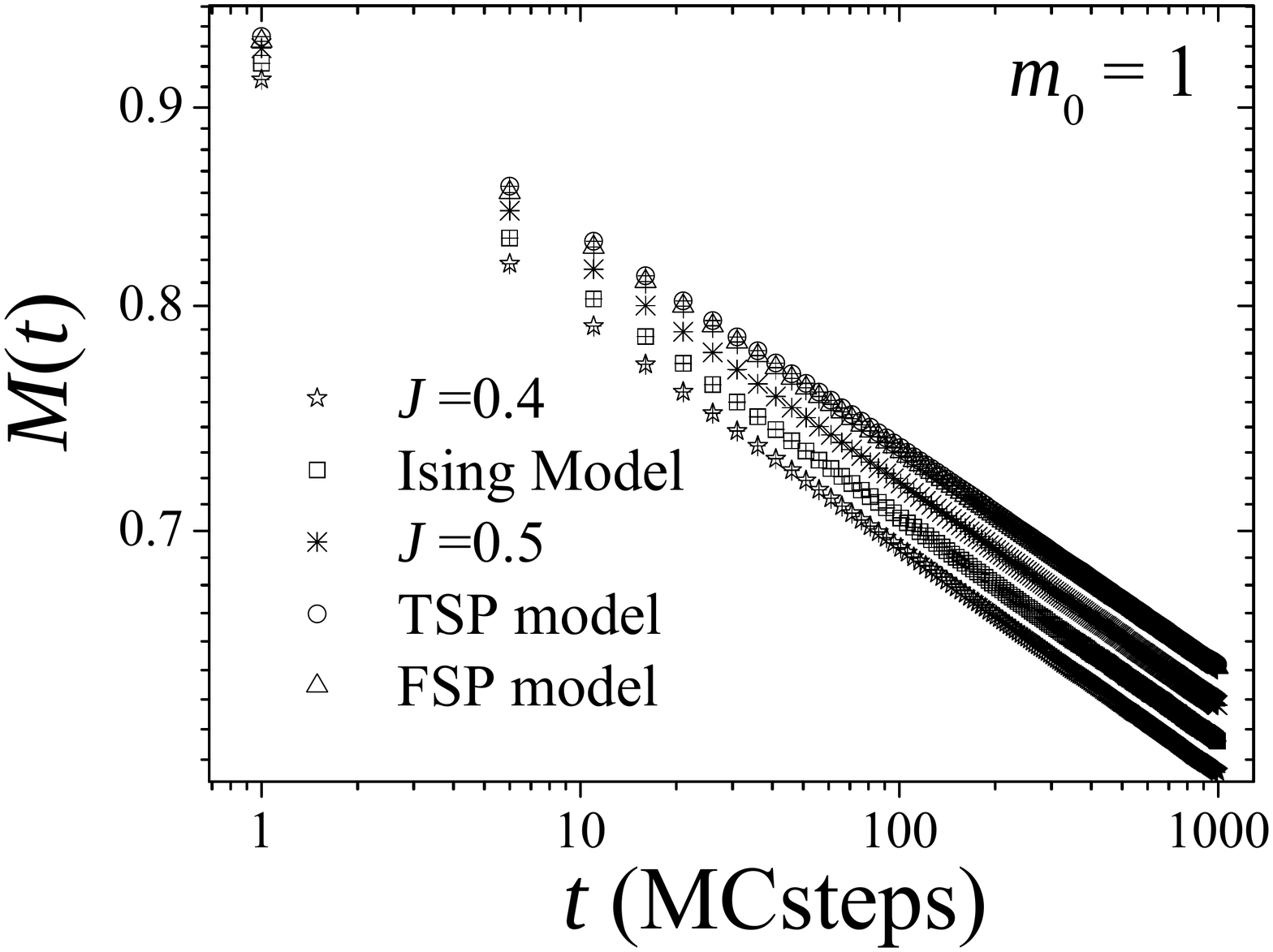} \ %
\includegraphics[width=\columnwidth]{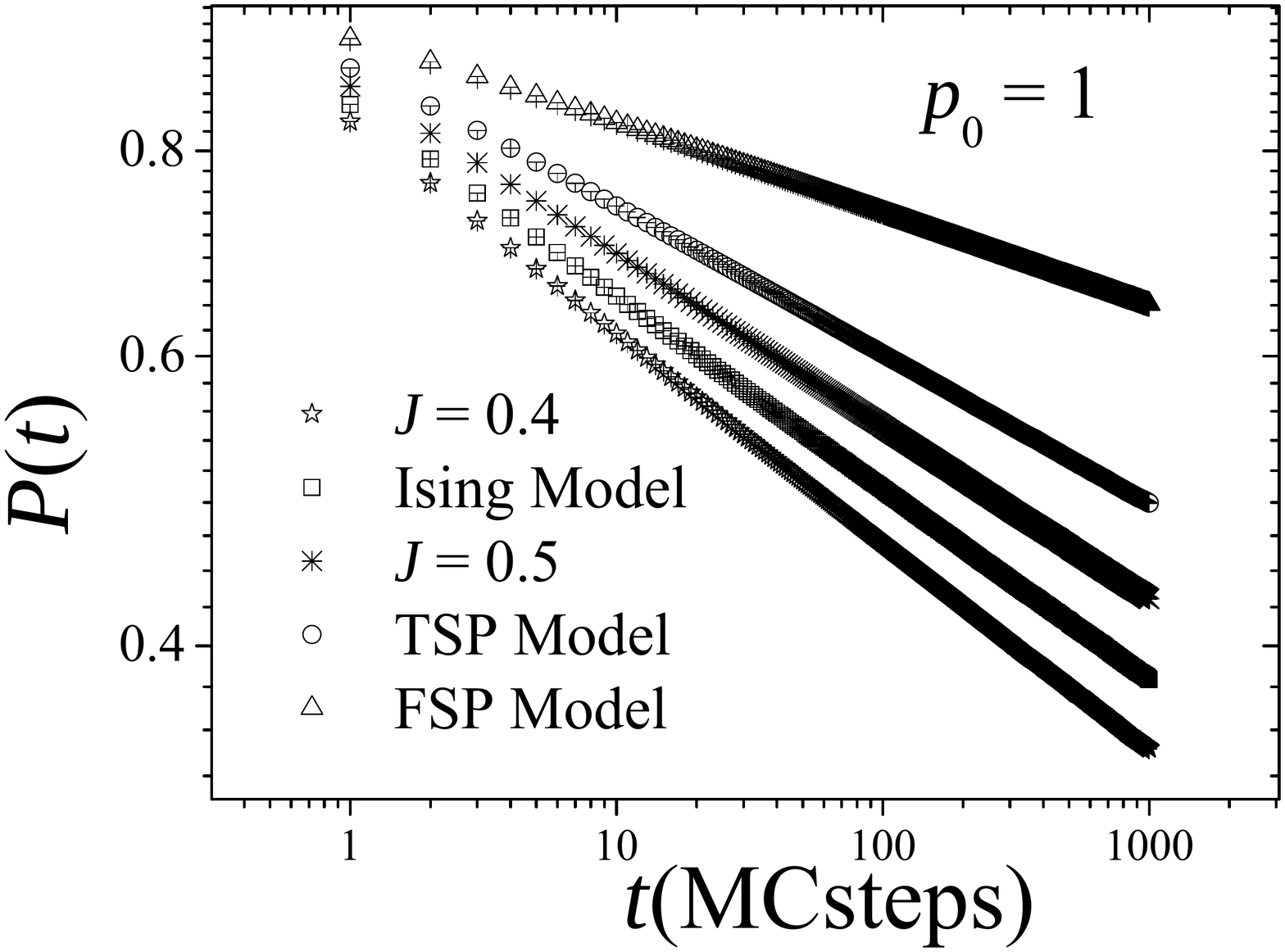}
\end{center}
\caption{Time evolving of $\overline{M}(t)$ (on the top) and $\overline{P}%
(t) $ (on the bottom), for $m_{0}=1\ $(or $p_{0}=1$) for the five coupling
constants.}
\label{Fig:time_decay_mag_pol}
\end{figure}

Here we proceed exactly as before to calculate $\nu $. We analyze the
external (over different time lags) and internal (over different bins)
variations to estimate the exponent $\mu =\beta /(\nu z)$. After a final
estimate of $\mu $ and with the previous estimates of $\nu $ and $z$, we
obtain an uncertainty for $\beta $: $\sigma _{\beta }^{2}=\beta ^{2}\left[
\left( \frac{\sigma _{\mu }}{\mu }\right) ^{2}+\left( \frac{\sigma _{\nu }}{%
\nu }\right) ^{2}+\left( \frac{\sigma _{z}}{z}\right) ^{2}\right] $. We
present our estimates of $\beta $ in Table \ref{Table:exponents_beta} as we
did for $\nu $ in Table \ref{Table:exponents_nu}.

\begin{table*}[tbh]
\centering%
\begin{tabular}{lllllllllll}
\hline
$J$ & $\beta _{m}$ & $\beta _{m}^{(boot)}$ & $\beta _{m}^{(\min )}$ & $\beta
_{m}^{(\max )}$ & $\beta _{p}$ & $\beta _{p}^{(boot)}$ & $\beta _{p}^{(\min
)}$ & $\beta _{p}^{(\max )}$ & $\beta _{m}^{\ast }$ & $\beta _{p}^{\ast }$
\\ \hline
0.4 & 0.1325(19) & 0.1326(24) & 0.1262(24) & 0.1409(49) & 0.2820(43) & 
0.2820(43) & 0.2711(31) & 0.2990(12) & $0.1360930$ & $0.2943721$ \\ 
Ising model & 0.1246(15) & 0.1241(20) & 0.1192(58) & 0.1310(30) & 0.2504(31)
& 0.2504(31) & 0.2404(56) & 0.2660(57) & $0.1250000$ & $0.2500000$ \\ 
0.5 & 0.1114(15) & 0.1100(18) & 0.1039(58) & 0.1173(40) & 0.1996(30) & 
0.1996(28) & 0.1927(43) & 0.2074(75) & $0.1143817$ & $0.2075268$ \\ 
TSP model & 0.1045(11) & 0.1046(15) & 0.1024(15) & 0.1092(33) & 0.1663(22) & 
0.1663(21) & 0.1566(55) & 0.1766(50) & $0.1041667$ & $0.1666667$ \\ 
FSP model & 0.08554(81) & 0.0853(11) & 0.0825(13) & 0.0895(17) & 0.0856(10)
& 0.0858(11) & 0.0814(43) & 0.0900(50) & $0.08333334$ & $0.0833334$ \\ \hline
\end{tabular}%
\caption{The static critical exponents $\protect\beta $ for the five
considered points, for magnetization ($m$) and polarization ($p$).\ All
estimates were obtained for the largest lattice used in this work: $L=256$.
Differently from $\protect\nu $ the conjectured values $\protect\beta ^{\ast
}$ for the magnetization and polarization are different.}
\label{Table:exponents_beta}
\end{table*}

Differently from what happens for $\nu $ (Table \ref{Table:exponents_nu}),
the conjectured values of $\beta $, for magnetization and polarization, are
different and our simulations corroborate both values. It is important to
notice that the exponents are the same for the FSP point. In order to test
the consistence of the estimates for

$\beta $ and $\nu $ we can compare $\beta /\nu $ with conjectured values
(see, for example, Ref. \cite{AlcarazDrugo}). It is important to stress that 
$\beta $ and $\nu $ may not be the same used in other papers and a
comparison must be done with some care.

\begin{figure}[h]
\begin{center}
\includegraphics[width=\columnwidth]{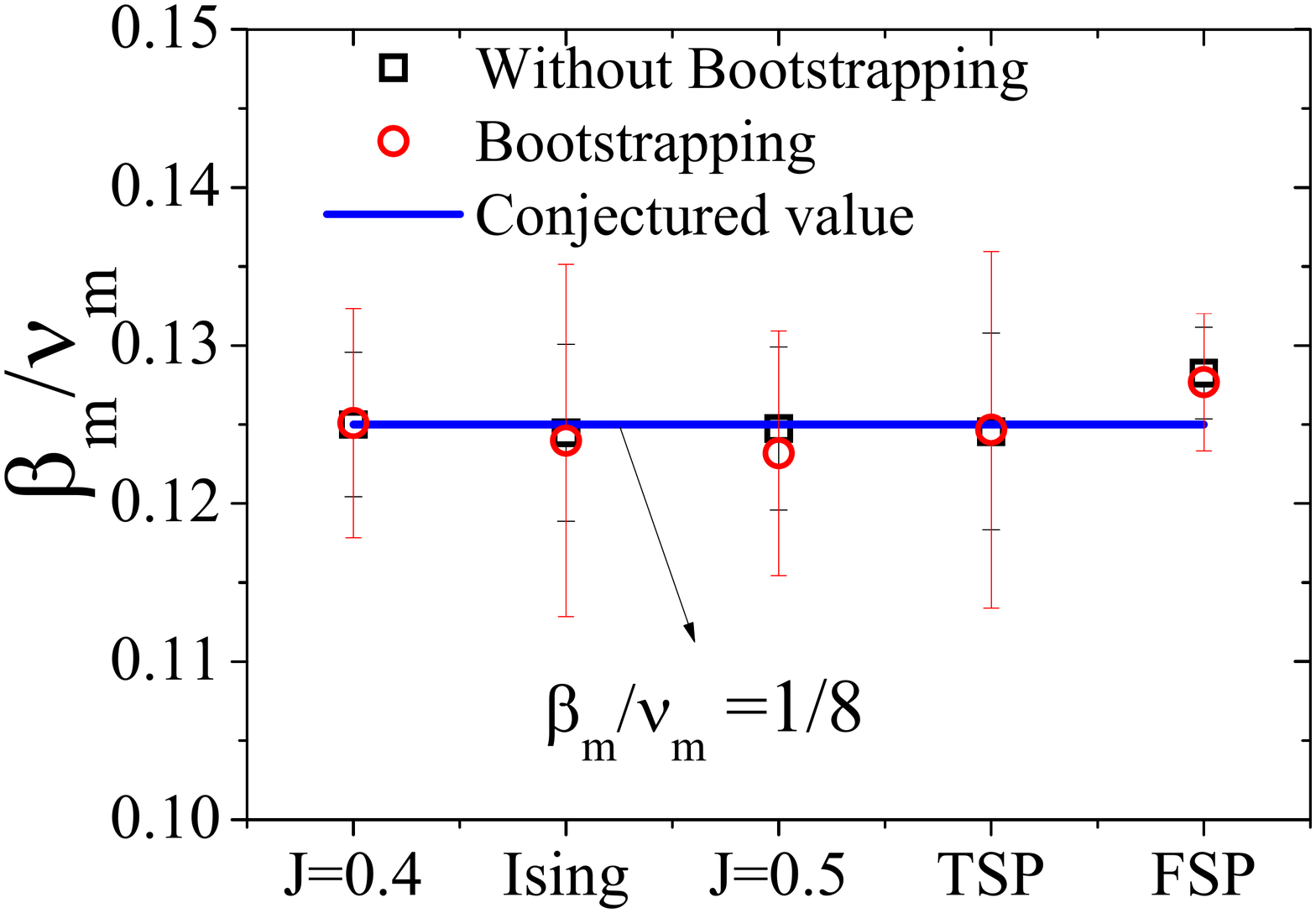} \ %
\includegraphics[width=\columnwidth]{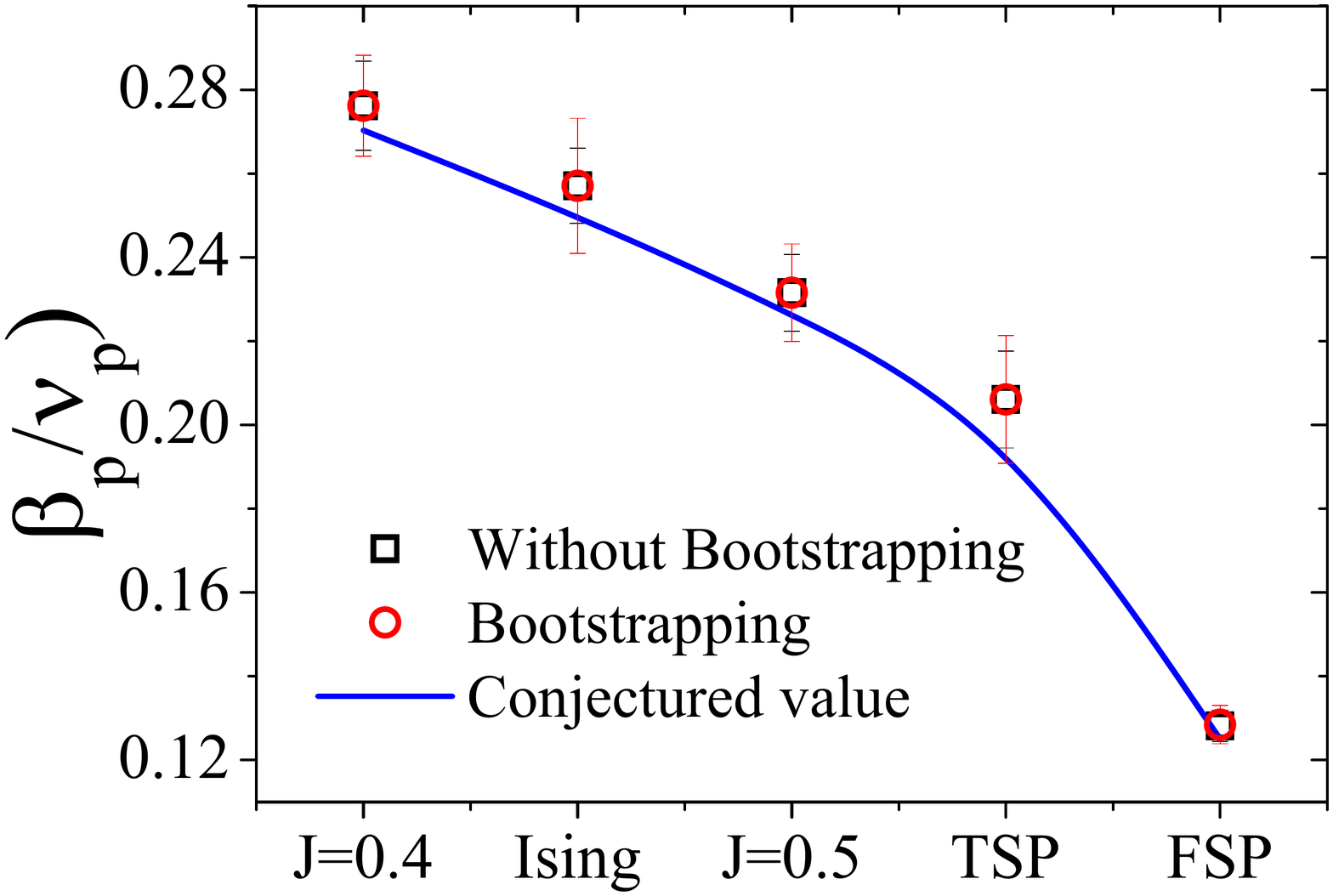}
\end{center}
\caption{The ratio $\protect\beta /\protect\nu $ calculate for magnetization
(on the top) and polarization (on the bottom). Our values present an
excelent agreement with the conjecture. }
\label{Fig:Beta_sobre_ni}
\end{figure}

Fig. \ref{Fig:Beta_sobre_ni} shows the ratio $\beta /\nu $ for the different
points. We can check that $\beta /\nu $ remains the same for all points in
the case of magnetization (on the top of this figure), while we have a
decrease of this ratio when $J$ increases for the polarization. In both
situations, an agreement with the conjectured values can be observed. The
blue curve was obtained using splines with the five conjectured points
obtained from literature. So we can check that all exponents and, moreover,
the conjectures are in agreement with our time-dependent Monte Carlo
simulations, i.e., the exponents can be obtained even out of equilibrium
extending even more the applicability of this wide and successful approach.

\section{Conclusion}

In this paper, we studied the non-equilibrium critical behavior of the
Ashkin-Teller model by performing Monte Carlo simulations far from
equilibrium. The dynamic critical exponents $\theta _{g}$, $\theta $, and $z 
$ were obtained for the two order parameters of the model: the magnetization
and polarization. The simulations were carried out on five different points
on the self-dual critical line including the Ising, $q=3$ and $q=4$ Potts
critical points. The exponents obtained on these points, for the
magnetization, are in good agreement when compared with available values in
the literature except for the exponent $z_{m}$ for the $q=4$ Potts critical
point that is slightly larger than that found in literature as well as the
exponent $\theta _{g_{m}}$ for the $q=3$ Potts critical point. We also found
for this critical point an exponent $\theta _{m}$ larger than those
presented in literature. Even with the Chatelain's argument \cite%
{Chatelain2004}, we think that further study is needed to explain this
difference in the exponents $\theta _{g_{m}}$ and $\theta _{m}$ of the
model. Besides, as stated by Li \textit{et al.} \cite{Li1997} and Takano 
\cite{Takano1996}, when studying the Ashkin-Teller and Baxter models,
respectively, the exponents $z_{m}$ and $z_{p}$ are almost constant but for
the $q=4$ Potts critical point.

We also obtained the static exponents. Our results after a careful method to
obtain the exponents presented a good agreement with conjectured results
from literature. The ratio $\beta /\nu $ decreases when $J$ increases for
the polarization but remains the same (according to our error bars) for the
magnetization.

In this work, we showed again the wide applicability of the theory of short
time dynamics to describe critical phenomena retrieving equilibrium
parameters in simulations out of equilibrium as well as predicting
nonequilibrium critical indexes. As an important additional contribution, we
also proposed a statistical approach to estimate exponents in time-dependent
MC simulations by composing fluctuations from intra and inter time-lags to
produce suitable error bars.

\subsection*{Acknowledgments}

This research work was in part supported financially by CNPq (National
Council for Scientific and Technological Development). R. da Silva would
like to thank Prof. L.G. Brunnet (IF-UFRGS) for kindly providing the
computational resources from Clustered Computing (ada.if.ufrgs.br)

\end{document}